\documentclass[twocolumn,showpacs,preprintnumbers,amsmath,amssymb]{revtex4}
\usepackage{graphicx}
\usepackage{dcolumn}
\usepackage{bm}

\begin{document}

\renewcommand{\theequation}{\arabic{section}.\arabic{equation}}

\title{Phase structure of $Z_2$ gauge theories for \\
frustrated antiferromagnets in two dimensions}

\author{Kazuya Nakane} 
\author{Akihiro Shimizu}
\author{Ikuo Ichinose}

\affiliation{Department of Applied Physics, Graduate School of 
Engineering, \\
Nagoya Institute of Technology, 
Nagoya, 466-8555 Japan 
}

\date{\today}

\begin{abstract}
In this paper, we study phase structure of $Z_2$ lattice gauge theories
that appear as an effective field theory describing 
low-energy properties of frustrated antiferromagnets in two dimensions.
Spin operators are expressed in terms of Schwinger bosons, 
and an emergent U(1) gauge symmetry reduces to a $Z_2$ gauge symmetry
as a result of condensation of a bilinear operator of the Schwinger 
boson describing a short-range spiral order.
We investigated the phase structure of the gauge theories
by means of the Monte-Carlo simulations,
and found that there exist three phases, phase with a long-range spiral order,
a dimer state, and a spin liquid with deconfined spinons.
Detailed phase structure and properties of phase transitions
depend on details of the models.

\end{abstract}

\maketitle

\section{Introduction}

In the last few decades, strongly-correlated electron systems
are one of the most intensively studied areas in the condensed
matter physics.
One may expect that some exotic phase appears as a result of the
interplay of strong correlations and quantum fluctuations.
Concerning to the high-$T_c$ cuprates, understanding of the under-doped regime
is still controversial.
Conventional Fermi-liquid picture may not hold in that region\cite{LNW}. 

Another intensively studied system is quantum magnets with
frustrations.
Study of that system has long history but its interests recently
revived because very interesting  experiments on the new materials 
like the organic Mott insulators $\kappa$-(ET)$_2$Z (Z=Cu[N(CN)$_2$]Cl,
etc)\cite{AF1} and X[Pd(dmit)$_2$]$_2$ 
(X=Me$_4$P, etc)\cite{AF2-1,AF2-2,AF2-3} have appeared.
Among them, the insulator with Z=Cu$_2$(CN)$_3$ has no long-range
order at low temperature\cite{AF3-1,AF3-2} and it is expected that 
a new type of
spin liquid, so called $Z_2$ spin liquid, is realized there\cite{Qi}.
Another interesting anisotropic triangular antiferromagnet is
Cs$_2$CuCl$_4$.
By neutron scattering, its spinon-like behaviors were 
observed\cite{CCC1,CCC2}.

To study possibility of exotic states in frustrated 
antiferromagnets like the $Z_2$ spin liquid, most studies employ
the Schwinger-boson representation for quantum spin operator.
As a result, there appear a local U(1) gauge symmetry and also an 
emergent gauge field.
Dynamics of the emergent gauge field strongly influences the 
structure of the ground state and low-energy excitations.
In the $Z_2$ spin-liquid scenario, the U(1) gauge symmetry is reduced to
a $Z_2$ symmetry because of appearance of a short-range spin spiral order, and 
$s={1 \over 2}$ spinons are deconfined and appear as a low-energy 
excitation\cite{Sachdev1}.
In order to obtain a conclusive proof of the existence of 
the $Z_2$ spin-liquid, reliable investigation on the gauge dynamics is 
necessary.
In the present paper, we shall report results of study on the 
$Z_2$ gauge theories obtained mostly by means of the Monte-Carlo (MC)
simulations.

The present paper is organized as follows.
In Sec.II, we shall introduce models of frustrated antiferromagnets
and review the Schwinger-boson representation of them.
We show that their low-energy effective model is a CP$^1$ gauge
model coupled with an additional doubly-charged vector field describing
a short-range spiral order.
In Sec.III, we shall show the phase structure of various effective
gauge models with local $Z_2$ gauge symmetry.
To obtain the phase diagrams, we calculated ``internal energy",
``specific heat", spin correlation functions and instanton density
by means of MC simulations.
There are three phases, phase with long-range order, dimer phase,
and spin liquid with deconfined spinons.
Section IV is devoted for conclusion and discussion.

\section{Frustrated antiferromagntes, Schwinger boson
and effective gauge theory}

\subsection{AF magnets and CP$^1$ gauge field theory}
\begin{figure}[ht]
\begin{center}
\includegraphics[width=6cm]{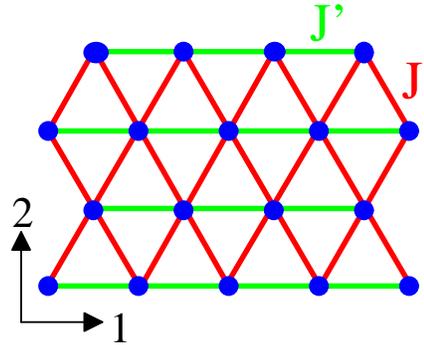}
\end{center}
\caption{Triangular lattice on which the Heisenberg model (\ref{Hspin})
is defined.}
\label{tri-lattice}
\end{figure}

Let us start with some specific model of a frustrated antiferromagnet
on the triangular lattice shown in Fig.\ref{tri-lattice}.
Exchange coupling in the horizontal bond is $J'$ and the 
others are $J$.
Quantum Hamiltonian ${\cal H}$ is given as
\begin{equation}
{\cal H}=J\sum \vec{S}_i\cdot\vec{S}_j
+J'\sum \vec{S}_i\cdot\vec{S}_j
+\cdots,
\label{Hspin}
\end{equation}
where $\vec{S}_i$ is $s={1 \over 2}$ spin operator at site $i$, 
and the ellipsis
denotes multi-spin and/or long-range interactions between spins, and 
the other notations are self-evident.

In the limit $J'/J \ll 1$, the system reduces to the usual antiferromagnets
on the square lattice and the ground state is expected to have 
the N\'eel order, whereas for $J'/J \sim 1$, a new state is expected to appear.
In order to study the system (\ref{Hspin}) by field-theory methods,
we introduce the Schwinger boson operators 
$a_i=(a_{\uparrow i},a_{\downarrow i})$ at each site $i$,
and then $\vec{S}_i$ is expressed as 
\begin{equation}
\vec{S}_i={1 \over 2}a^\dagger_i \vec{\sigma}a_i,
\label{SS}
\end{equation}
where $\vec{\sigma}$ are the Pauli spin matrices.
The following local constraint must be imposed as the physical-state
condition in the Schwinger boson Hilbert space,
\begin{equation}
(a^\dagger_{\uparrow i}a_{\uparrow i}+a^\dagger_{\downarrow i}a_{\downarrow i})
|\mbox{Phys}\rangle=|\mbox{Phys}\rangle.
\label{const}
\end{equation}

We employ the path-integral methods to investigate the quantum system,
and introduce CP$^1$ variables 
$z_i=(z_{\uparrow i},z_{\downarrow i})=(z_{1 i},z_{2 i})$ 
corresponding to $a_i$, which satisfy the constraint
\begin{equation}
\bar{z}_{\uparrow i}z_{\uparrow i}+\bar{z}_{\downarrow i}z_{\downarrow i}
=1,
\label{CP1}
\end{equation}
at each site $i$ and $\bar{z}_{\uparrow, \downarrow, i}$ is the 
complex conjugate of ${z}_{\uparrow, \downarrow, i}$.
From the Hamiltonian (\ref{Hspin}), the partition function is given as 
\begin{equation}
Z=\int[D\bar{z}Dz]_{{\rm CP}^1}\exp \Big[-\int d\tau 
\Big(\sum_i\bar{z}_i\cdot\dot{z}_i+{\cal H}(\bar{z},z)\Big)\Big],
\label{Z}
\end{equation}
where $\tau$ is the imaginary time, $\dot{z}_i={dz_i \over d\tau}$
and $\int[D\bar{z}Dz]_{{\rm CP}^1}$ denotes the integration over
CP$^1$ variables $z_i$'s satisfying the constraint (\ref{CP1}).
${\cal H}(\bar{z},z)$ is derived from (\ref{Hspin}) and (\ref{SS}).
The above system is obviously invariant under a 
{\em local gauge transformation}
$z_i(\tau)\rightarrow e^{i\theta_i(\tau)}z_i(\tau)$ with an arbitrary
$\theta_i(\tau)$ satisfying $\theta_i(+\infty)=\theta_i(-\infty)$.

In the limit $J'\rightarrow 0$, an effective field theory is obtained from
the partition function $Z$ in (\ref{Z}) by integrating out the 
high-energy modes of $z_i$ (or $z_i$'s on all odd sites\cite{CP-1,CP-2}).
The resultant theory is a CP$^1$ gauge model, which is described by
the following action $S_z$ in the continuum 
spacetime with coordinate $x_{\mu}=(x_0=\tau, x_1,x_2)$,
\begin{equation}
S_z=\int d^3x\Big[{1 \over g^2}\sum_\mu |D_\mu z|^2
+{1 \over e^2}\sum_{\mu<\nu}F_{\mu\nu}^2\Big],
\label{Sz}
\end{equation}
where $D_\mu z=(\partial_\mu+iA_\mu)z$, 
$F_{\mu\nu}=\partial_\mu A_\nu-\partial_\nu A_\mu$
with emergent gauge field $A_\mu$.
In Eq.(\ref{Sz}), $g,\; e$ are coupling constants.
Bare value of $g$ is independent of the antiferromagnetic
(AF) exchange coupling $J$,
but it measures the solidity of the AF order, i.e., additional 
interactions that enhance (suppress) the AF order decrease (increase) 
the value of $g$.  
On the other hand, the bare value of $1/e$ is vanishing
for the AF Heisenberg model with only the nearest-neighbor
(NN) coupling but it acquire
a finite value due to the renormalization effect of the high-energy
modes.
Multi-spin nonlocal interactions like a ring exchange coupling generate 
nonvanishing value of $1/e$\cite{Sawa}.
Varying the parameters $g$ and $e$ induces a phase transition and
the structure of the ground state and low-energy excitations change
drastically through the phase transition as we see in the following
sections.

The field theory defined by (\ref{Sz}) is obviously invariant
under a U(1) gauge transformation.
The continuum description (\ref{Sz}) makes it unclear if this
U(1) gauge symmetry is compact or noncompact one.
As the original system of the AF magnets is defined on the lattice
and transformation parameter $\theta_i(\tau)$ is defined mod $2\pi$, 
one may expect that the model (\ref{Sz}) is a compact U(1) gauge
system, in which topological nontrivial objects like instantons
and vortices can exist.
This expectation is qualitatively correct, but contribution 
from instanton configurations to the partition function 
is partly suppressed if there exists a Berry-phase term, 
$\int d^3x \epsilon_{\mu\nu\lambda}\partial_\mu F_{\nu\lambda}$
(where $\epsilon_{\mu\nu\lambda}$ is the antisymmetric tensor), 
in the action in addition to $S_z$\cite{Berry-1,Berry-2,Berry-3}.
For the case $J' \neq 0$, it is not easy to calculate the coefficient of
the Berry phase, which plays a crucial role in the suppression of 
instantons.
We shall not consider its effect in the following numerical investigation, 
and give comments on it in Sec.IV\cite{CS}.

\begin{figure}[ht]
\begin{center}
\includegraphics[width=7cm]{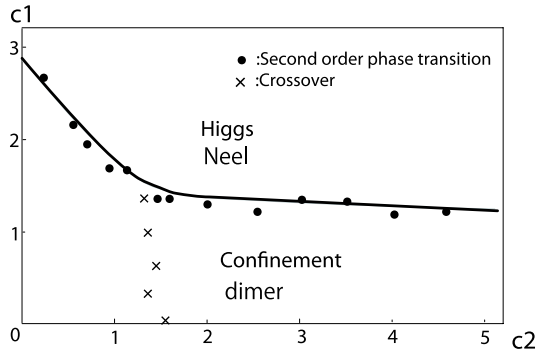}
\end{center}
\caption{Phase diagram of the U(1) gauge theory of CP$^1$ spinons\cite{CPN-2}.
There are two phases.}
\label{cp1u1_phase}
\end{figure}

Phase structure of the CP$^{N-1}$ field theory has been
studied by the $1/N$-expansion and numerical methods\cite{CPN-1,CPN-2,CPN-3}.
For the compact U(1) gauge case, a lattice-regularized version
of (\ref{Sz}) is quite useful for investigation on 
the CP$^1$ gauge model, and its action is given as follows,
\begin{eqnarray}
A_z &=& {c_1 \over 2}\sum_{x,\mu}\bar{z}_{x+\mu}U_{x,\mu}z_x
+{c_2 \over 2}\sum_{x,\mu<\nu}U_{x,\mu}U_{x+\mu,\nu}
\bar{U}_{x+\nu,\mu}\bar{U}_{x,\nu}  \nonumber \\
&&+\mbox{c.c.},
\label{Az}
\end{eqnarray}
where $x$ denotes site of the {\em cubic lattice}, and  
the coupling $c_1$ corresponds to $1/g^2$ and 
$c_2$ to $1/e^2$.
Phase diagram has been obtained in the $c_1-c_2$ plane. 
See Fig.\ref{cp1u1_phase}.
There are two phases separated by the critical line $c_1=c_{1c}(c_2)$, 
one of which corresponds to the N\'eel state
for $c_1>c_{1c}(c_2)$ and the other state is a dimer state 
$c_1<c_{1c}(c_2)$ in which the spinon $z_x$ is confined to a 
spin-triplet excitation $\bar{z}_x\vec{\sigma}z_x$.
The phase transition across the transition line is 
of second order, and it belongs to the universality class of the $O(3)$
nonlinear sigma model in three dimensions ($3D$) 
(for small to medium values of $c_2$).
The $s={1 \over 2}$ AF Heisenberg model corresponds to
$c_1>c_{1c}$, and the ground state has the AF long-range order.
By introducing an {\em inhomogeneity} in the exchange coupling $J$
that enhances dimerization, 
the value of $c_1$ in the effective model (\ref{Az}) is decreased and the 
phase transition takes place from the N\'eel to dimer states\cite{Yoshi}.
Recently, numerical study on the inhomogeneous SU(2) AF Heisenberg model,
which is essentially the same with that studied in Ref.\cite{Yoshi},
was performed quite in detail and  
the existence of the phase transition from the N\'eel to dimer states
was verified\cite{Janke1}.
Phase transition belongs to the universality class of the $3D$ $O(3)$ 
nonlinear-sigma model,
as predicted by the study of the effective lattice model (\ref{Az}).

As shown in Fig.\ref{cp1u1_phase}, the deconfined Coulomb phase does 
not exist in the 
model (\ref{Az}) of the U(1) gauge theory.
Appearance of the Coulomb phase requires long-range and nonlocal
interaction of gauge field $U_{x\mu}$,
which may be generated by the coupling with gapless 
fermions\cite{nonlocal,nonlocal2}.
In the pure quantum spin models without doping of holes,
the deconfined phase is expected to appear by introducing frustrations
because the Higgs mechanism is expected to take place by
the appearance of the (short-range) spiral order.
In that case, the U(1) gauge symmetry spontaneously breaks down to $Z_2$.
It is known that the deconfined phase exists in the $3D$ $Z_2$ gauge models.
There are interesting studies on spin liquids with deconfined spinons
in the framework of the $Z_2$ gauge model.
However, detailed and reliable study on the phase structure of the 
$Z_2$ gauge models relevant to the frustrated spin systems is still lacking.
We study this problem in this paper.

Before going into details of the study on the frustrated AF magnets,
let us comment on the validity of the present methods using the lattice
field theory for studying AF magnets.
To define quantum many-body systems without ambiguities, an 
ultra-violet (UV) regularization is necessary.
In quantum spin models like (\ref{Hspin}), the spatial lattice naturally
gives such an UV regularization.
In the present approach, we first study the original model carefully and 
identify the relevant modes in the low-energy and low-momentum
region.
Through these observations, we obtain an effective field theory in the
continuum spacetime.
Then in order to study the effective field theory nonperturbatively
(e.g. by means of the MC simulations), we reformulate it by using a
{\em spacetime} lattice as a systematic regularization.
Structure of the lattice model is deterimined by the symmetry 
of the effective field theory and we expect that details of the
lattice model does not influence substantially physical results like
phase structure and critical behaviors by the unversality-class argument.
For the quantum SU(2) AF magnets, it is known 
that the results obtained by the effective CP$^1$
lattice model (\ref{Az}) are in good agreement with those obtained for 
the original AF Heisenberg model, as we explained above.
Furthermore, phase structure of the lattice CP$^n \; (n=1,\cdots,4)$ models
obtained by the MC simulations is the same with that obtained by the 
$1/N$-expansion for the CP$^{N-1}$
field theory in the continuum spacetime\cite{CPN-1,CPN-2}. 
These facts encourage us to apply the same methods to more
complicated quantum spin systems like triangular AF spin systems
with frustrations.
More comments on the reliablity of the methods
will be given in Sec.IV, after showing the main results
of the present study in the following sections.


\subsection{Effect of frustrations}

The effect of the frustration in the AF magnets (\ref{Hspin}) can be studied 
in the framework of the CP$^1$ gauge field theory whose action
has the following term $S_\Lambda$ in addition to $S_z$\cite{Sachdev1},
\begin{eqnarray}
S_\Lambda &=&\int d^3x\sum_{\alpha=1,2}
\Big[{1 \over g^2_\Lambda} |D^{(2)}_\mu \Lambda_\alpha|^2
+m_\Lambda|\Lambda_\alpha|^2+\lambda |\Lambda_\alpha|^4 \nonumber \\
&&+i\Lambda_\alpha\bar{z}\partial_\alpha \tilde{z}+\mbox{c.c.}\Big],
\label{SL}
\end{eqnarray}
where $\Lambda_\alpha\;(\alpha=1,2)$ is a {\em doubly-charged spatial vector 
field},
$D^{(2)}_\mu=\partial_\mu+2iA_\mu$, and 
$\tilde{z}_a(x)=\epsilon_{ab}\bar{z}_b(x)$
($\epsilon_{12}=-\epsilon_{21}=1, \; \epsilon_{11}=\epsilon_{22}=0$).
Origin of the new term $S_\Lambda$ is as follows.
The $J'$-term in Eq.(\ref{Hspin}) generates terms like 
${J' \over J}\sum_{ij}|\bar{z}_i\cdot {\tilde{z}}_j|^2$ in the 
effective field theory, where the extra factor $1/J$ comes from
the redefinition of the imaginary time $\tau\rightarrow \tau\times (aJ)$
($a=$lattice spacing$=$often set unity).
After inserting the following identity into the path-integral representation
of the partition function,
\begin{equation}
\int d\Lambda_{ij}d\bar{\Lambda}_{ij}\;
e^{-{J' \over J}
({J \over J'}\Lambda_{ij}-iz_i\cdot\bar{\tilde{z}}_j)
({J \over J'}\bar{\Lambda}_{ij}+i\bar{z}_i\cdot\tilde{z}_j)}=\mbox{constant}
\label{Lambda}
\end{equation}
the above quartic term of $z_i$ is decoupled by a Hubbard-Stratonovich
field $\Lambda_\alpha$.
By the effects of renormalization of high-momentum modes,
the extra terms in $S_\Lambda$ and renormalization of the mass, 
which preserve the local U(1)
gauge symmetry, appear for describing low-energy behavior of the
system\cite{FNLambda}. 

Physical meaning of $S_\Lambda$ becomes transparent by considering the case
$m_\Lambda<0$.
In this case, we expect the nonvanishing expectation value of the field
$\Lambda_\alpha$, i.e., $\langle \Lambda_\alpha\rangle \neq 0$.
By solving the field equation derived from the action $S_z+S_\Lambda$,
it is straightforward to verify that the low-energy configurations are 
given by,
\begin{equation}
z_a(x)={1 \over \sqrt{2}}\Big(v_a(x)
e^{ig^2\langle \vec{\Lambda} \rangle\cdot
\vec{x}}+\epsilon_{ab}\bar{v}_b(x)
e^{-ig^2\langle \vec{\Lambda} \rangle\cdot\vec{x}}\Big),
\label{zv}
\end{equation}
where $v_a(x)\; (a=1,2)$ is a slowly varying complex 
field satisfying $\sum_a|v_a(x)|^2=1$\cite{FNgauge,FNU1}.
For the configurations given by (\ref{zv}), the SU(2) spin field 
$\vec{S}(x)\equiv \bar{z}(x)\vec{\sigma}z(x)$ has the
following form,
\begin{eqnarray}
&& \vec{S}(x)= \vec{n}_1\cos(2g^2\langle \vec{\Lambda} \rangle\cdot\vec{x})
+\vec{n}_2\sin(2g^2\langle \vec{\Lambda} \rangle\cdot\vec{x}),\nonumber \\
&& \vec{n}_1= {\rm Re}[\bar{\tilde{v}}\vec{\sigma}v], \;\;
\vec{n}_2={\rm Im}[\bar{\tilde{v}}\vec{\sigma}v], \nonumber  \\
&& \vec{n}^2_1=\vec{n}^2_2=1, \;\;
\vec{n}_1\cdot\vec{n}_2=0.
\label{spiral}
\end{eqnarray}
On the other hand, the ``spin-nematic field" 
$\vec{n}_3=\vec{n}_1\times \vec{n}_2$ is given as 
$\vec{n}_3=\bar{v}\vec{\sigma}v$.
It is obvious that $\vec{S}(x)$ in (\ref{spiral}) corresponds to
a spiral state if $\langle \vec{n}_1 \rangle \neq 0, \;
\langle \vec{n}_2 \rangle \neq 0$.

By substituting Eq.(\ref{zv}) and 
$\Lambda_{0\alpha}=\langle \Lambda_\alpha \rangle$ into 
the continuum action $S_z+S_\Lambda$, low-energy effective theory
is obtained.
Condensation of $\Lambda_\alpha$ not only generates the spiral 
state of $\vec{S}$ but also a finite mass of the gauge field $A_\mu$.
CP$^{N-1}$ model with a massive ``gauge field" has been studied in the
continuum spacetime by the $1/N$-expansion, 
but the obtained results are not reliable for
the case of finite $N$ (in particular the case $N=2$) because an
important effect at $O(1/N)$ coming from topological excitations is 
totally ignored there\cite{massiveCPN}.
In fact, the condensation of $\Lambda_\alpha$ preserves 
the local $Z_2$ gauge invariance of the system because it carries double
charge, and therefore the topological
nontrivial excitation carrying a half-magnetic quantum, dubbed vison, 
exists as a low-energy excitation\cite{vison}.
Also in Ref.\cite{CSS}, a quantum phase transition between a spin liquid
with deconfined spinons and magnetically orderd state was studied,
and various physical quantities were calculated by the $1/N$-expansion
in an effective CP$^{N-1}$ field theory with a global U(1) symmetry.
There it is assumed that effect of the vison can be ignored.
Our study of the gauge model with the full $Z_2$ gauge symmetry 
in the present paper will show that this assumption is correct.
See, for example, the calculation of the instanton density in Sec.III.

In the rest of the present paper, we shall study the effective $Z_2$
gauge theories obtained by substituting 
$\vec{\Lambda}=\langle \vec{\Lambda} \rangle$ and Eq.(\ref{zv})
into the action $S_z+S_\Lambda$.
To this end, we reformulate it by using
the lattice regularization that preserves the local $Z_2$ gauge symmetry.
We use a cubic spacetime lattice because frustrations coming from AF
coupling on the triangular lattice has disappeared by using 
the parameterization (\ref{zv}).
The resultant lattice model is explicitly given by the following action,
\begin{eqnarray}
A(c_3)&=&
{c_1 \over 2}\sum_{x,\mu}(\bar{v}_{x+\mu}U_{x,\mu}v_x
+\bar{v}_xU_{x,\mu}v_{x+\mu})  \nonumber \\
&&+{c_2 \over 2}\sum_{x,\mu<\nu}U_{x,\mu}U_{x+\mu,\nu}
\bar{U}_{x+\nu,\mu}\bar{U}_{x,\nu} \nonumber  \\
&&+{c_3 \over 2}\sum_{x,\mu}U^2_{x,\mu}+\mbox{c.c.},
\label{Z2action1}
\end{eqnarray}
where we explicitly show the dependence of the parameter $c_3$ 
in $A(c_3)$,
as we study the model with fixed values of $c_3$ in the following section.
From the above consideration, 
$c_3 \propto \langle \vec{\Lambda}\rangle ^2$.
Partition function of the gauge model (\ref{Z2action1}) is given as
\begin{equation}
Z_{\rm Gauge}=\int[D\bar{z}Dz]_{{\rm CP}^1}[D\bar{U}DU] \;
\exp A(c_3).
\label{Zgauge}
\end{equation}
It is obvious that the system (\ref{Z2action1}) has a local
$Z_2$ gauge symmetry instead of the U(1) symmetry.
Then we call $v_x$ $Z_2$CP$^1$ boson.
In the limit $c_3 \rightarrow \infty$, configurations of the gauge field 
are restricted to $U_{x,\mu}=\pm 1$ and the model reduces to a $Z_2$
gauge system.
In the case $c_3 \rightarrow \infty, \;c_1=0$, the system is 
the pure $Z_2$ gauge model in $3D$, which is dual to
the $3D$ Ising model and has a second-order phase transition 
from the confined to deconfined ``Coulomb" phases as $c_2$ is increased.
This is in sharp contrast to the U(1) gauge model in $3D$, in which only
the confined phase exists.
As the deconfined phase corresponds to spin liquid with weakly 
interacting spinons, one may expect realization of a fractionalization 
phenomenon in frustrated AF magnets.
In the following sections, we shall study phase structure of the  
model (\ref{Z2action1}) by means of the MC simulations.

\section{Numerical studies}
\begin{figure}[th]
\begin{center}
\includegraphics[width=8cm]{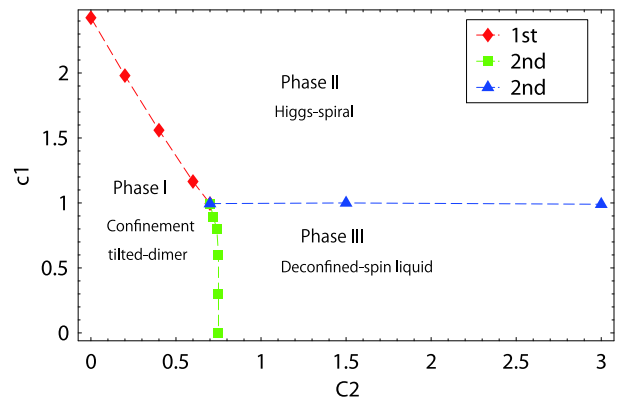}
\end{center}
\caption{Phase diagram of the $Z_2$ gauge theory of CP$^1$ spinons.
There are three phases.}
\label{phase-z2}
\end{figure}
\begin{figure}[h]
\begin{center}
\includegraphics[width=7cm]{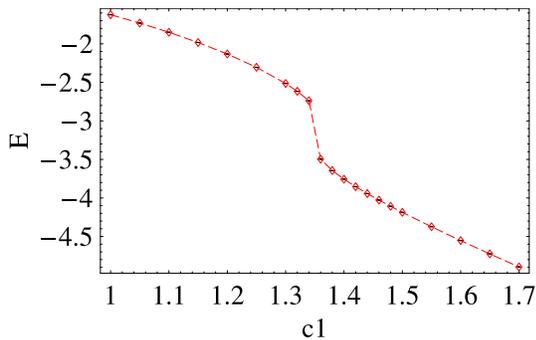}
\end{center}
\caption{$E$ for $c_2=0.5$. There is sharp discontinuity at $c_1\simeq 1.35$
that indicates a first-order phase transition. System size is $L=24$.}
\label{E-z2c205}
\end{figure}
\subsection{$Z_2$ lattice gauge model of $Z_2$CP$^1$ spinon}
We first study the $Z_2$ gauge model coupled with the field $v_x$
that corresponds to the limit $c_3 \rightarrow \infty$ of the model
(\ref{Z2action1}).
It is known that $Z_2$ gauge model coupled with single component Higgs boson 
describes nematic phase transition, and its phase structure was studied
by both analytical and numerical methods\cite{nematic}.
In these studies, importance of topological line defects 
(world lines of vison) was emphasized.

In order to investigate the phase structure of the $Z_2$ gauge model,
we defined the model on the cubic lattice of size $L^3$ with the
periodic boundary condition and 
calculated the ``internal energy" $E=\langle A(\infty) \rangle/L^3$,
the ``specific heat" $C=\langle (A(\infty)-E)^2\rangle/L^3$, etc.
We used the standard Metropolis algorithm  for the MC simulations\cite{MC}.
The typical statistics used was $10^5$ MC steps for each sample,
and the averages and errors were estimated over $10\sim 20$ samples.
Average acceptance probability was about $40\sim 50\%$.

The obtained phase diagram is shown in Fig.\ref{phase-z2}.
There are three phases, and calculation of various physical
quantities gives the following identifications.
\begin{enumerate}
\item In the phase I, there is no AF long-range
order and the gauge dynamics is realized in the confined phase.
Low-energy excitations are spin-triplet bound states of the spinor
$v_x$ (triplon), i.e., $\vec{n}_1(x), \; \vec{n}_2(x)$ in Eq.(\ref{spiral}).
We call this phase tilted dimer state.
\item In the phase II, there exists the magnetic long-range order of $v_x$,
which corresponds to the spiral order of $\vec{S}(x)$, i.e., 
$\langle \vec{n}_1\rangle\neq 0,\; \langle \vec{n}_2\rangle\neq 0$.
The gauge dynamics is in the Higgs phase because of the condensation of $v_x$.
Low-energy excitations are gapless spin wave described by 
{\em uncondensed component} of $v_x$.
\item Phase III represents the paramagnetic spin liquid state. 
As for gauge dynamics, a deconfined ``Coulomb phase" is realized, and
the number of topological vortices is conserved.
Low-energy excitations are massive spinon $v_x$.
\end{enumerate}

\begin{figure}[t]
\begin{center}
\includegraphics[width=5cm]{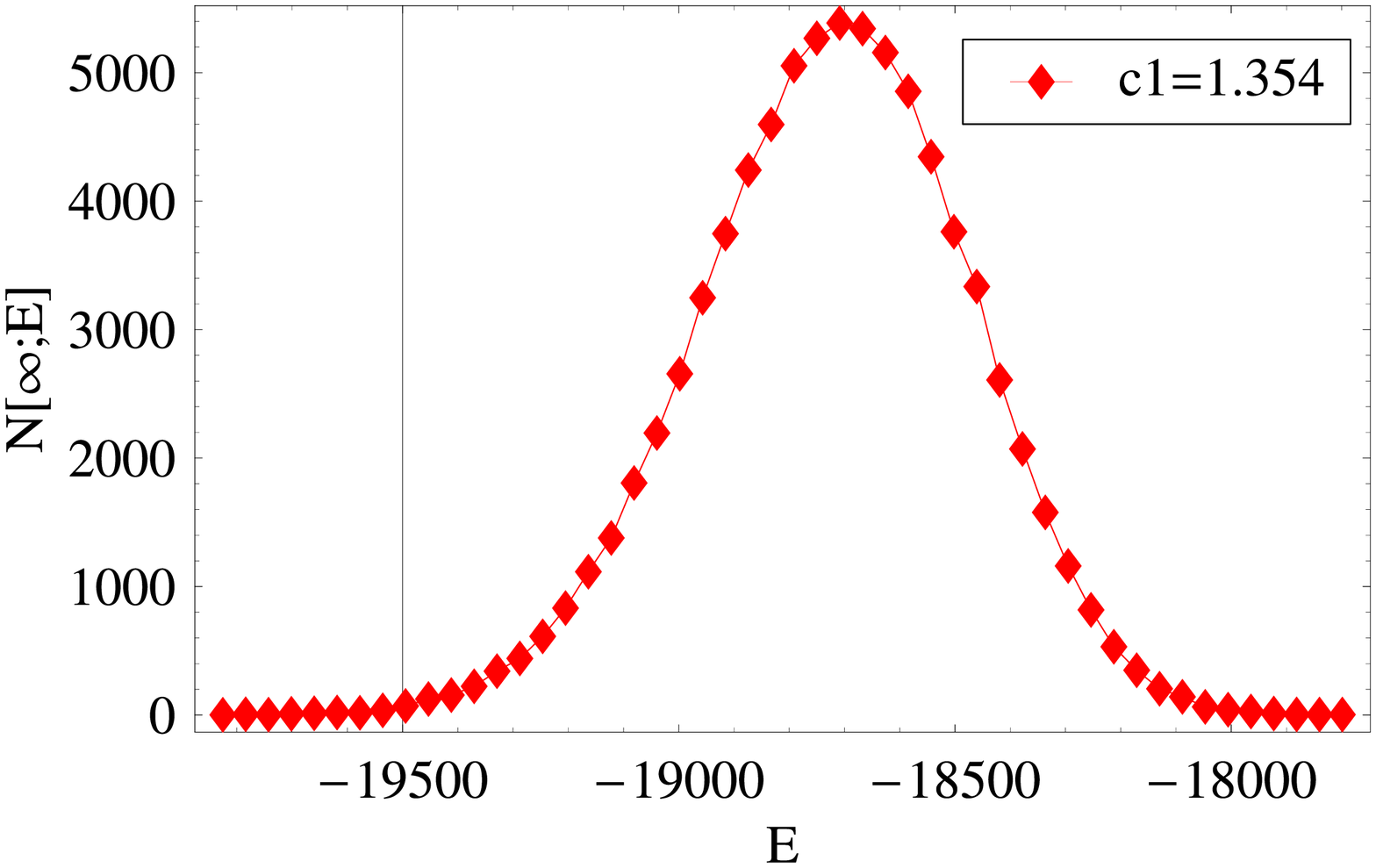}
\includegraphics[width=5cm]{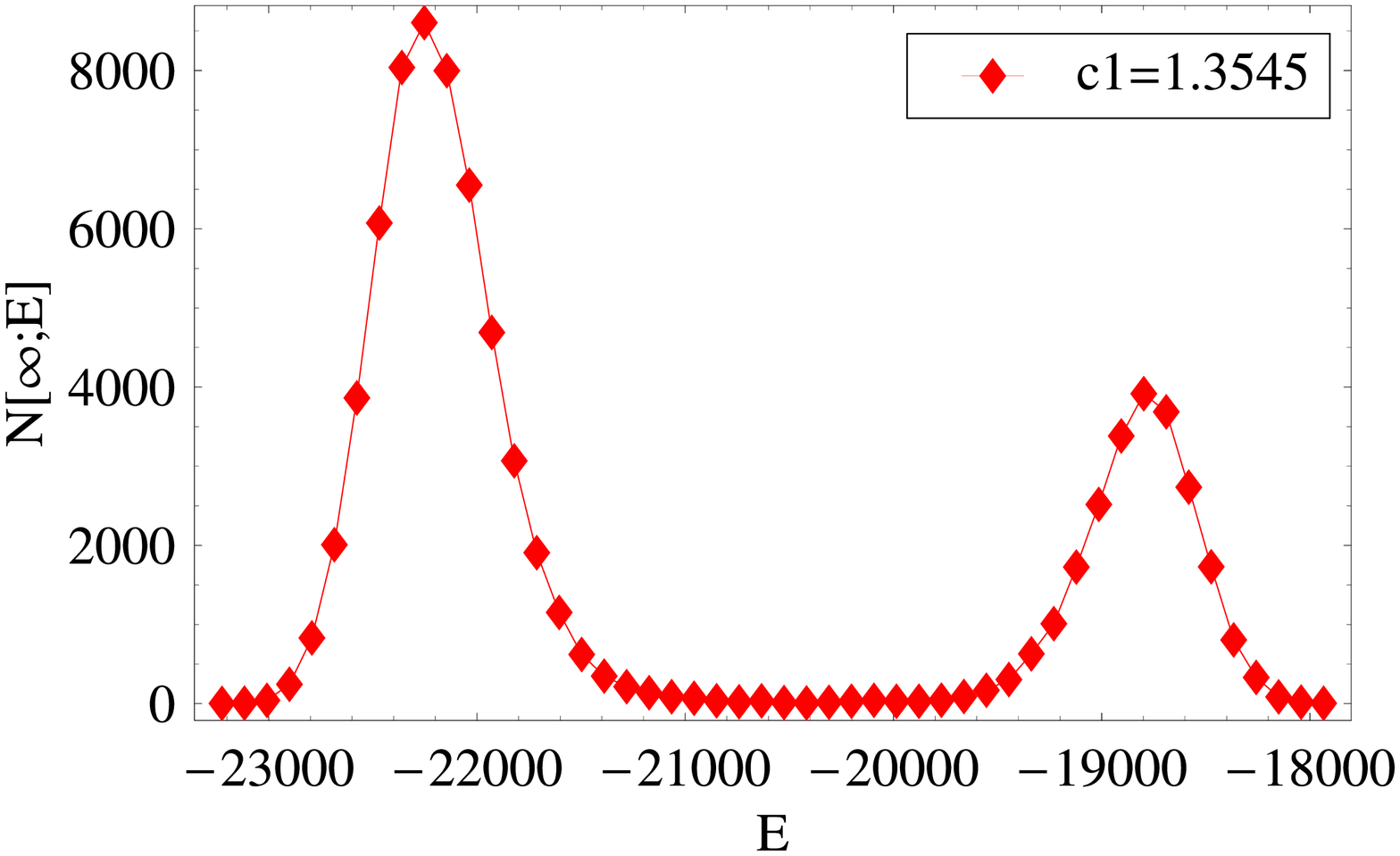}
\includegraphics[width=5cm]{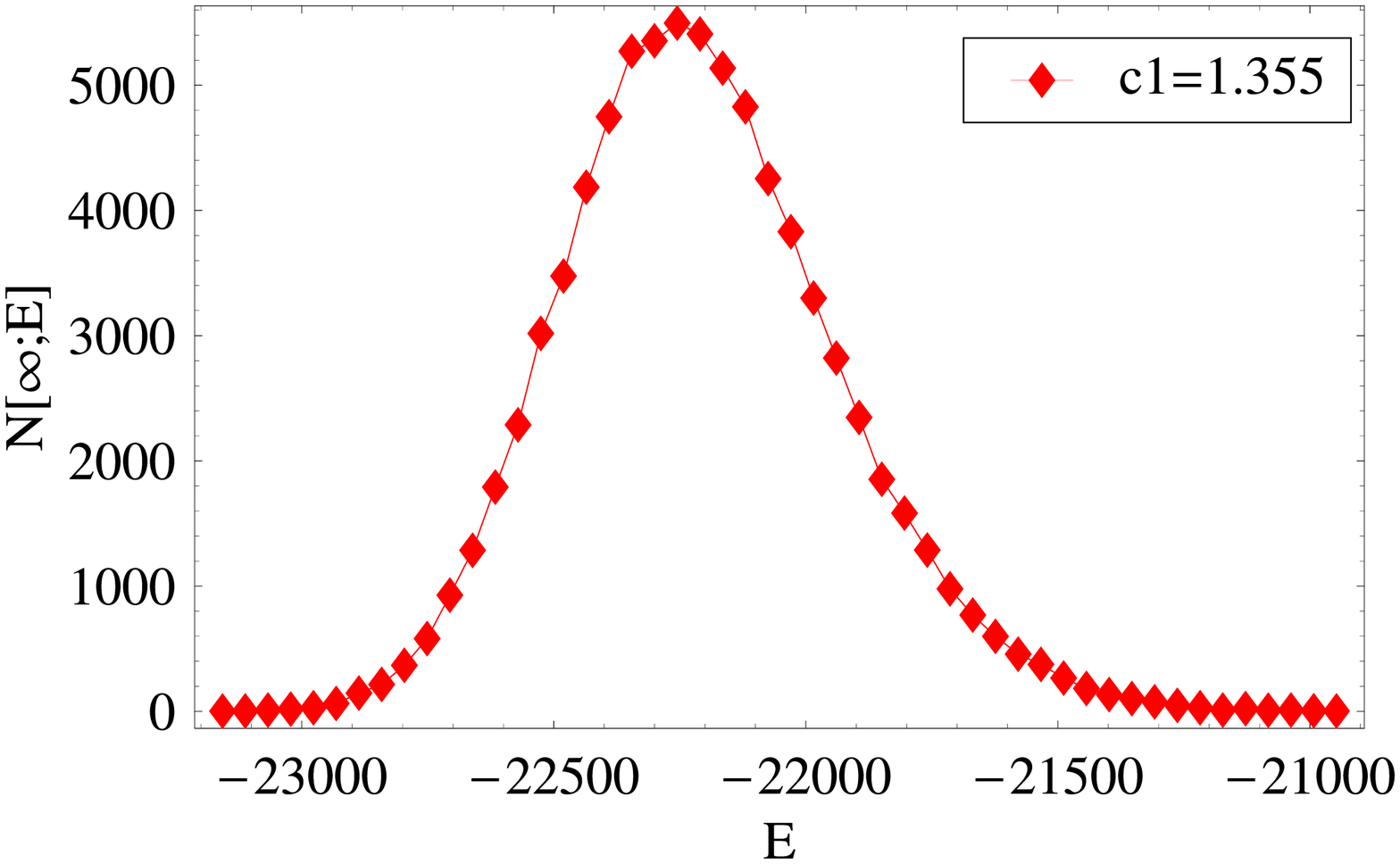}
\end{center}
\caption{Distribution of $A(\infty)$ for $c_2=0.5$ and $c_1$ close to
the phase transition point. The double-peak structure at $c_1=1.3545$ 
confirms the existence of the first-order phase transition.
System size $L=24$.}
\label{hist-z2c205}
\end{figure}

We first show the numerical calculations of $E$ and $C$ for
establishing the above phase diagram in Fig.\ref{phase-z2}.
We first focus on the transition from phase I to II.
In Fig.\ref{E-z2c205}, we show calculation of $E$ as a function of $c_1$
with $c_2=0.5$.
It is obvious that there exists a sharp discontinuity at $c_1\simeq 1.35$,
which indicates a first-order phase transition.
In order to verify it, we measured distribution of values of $A(\infty)$
generated in the MC steps, ${\cal N}[\infty; E]$, which is defined as 
\begin{eqnarray}
Z_{\rm Gauge}&=&\int dE 
\int[D\bar{z}Dz]_{{\rm CP}^1}[D\bar{U}DU] \;
e^{A(\infty)}\delta(A(\infty)-E)  \nonumber  \\
&=&\int dE \; {\cal N}[\infty; E].
\label{NA}
\end{eqnarray}
We show the result near the critical point in Fig.\ref{hist-z2c205}.
At $c_1=1.3545$, ${\cal N}[\infty;E]$ has a double-peak structure,
whereas the others have a single peak.
From this result, we judge that the first-order phase transition
takes place at $c_1=1.3545$.

\begin{figure}[th]
\begin{center}
\includegraphics[width=7cm]{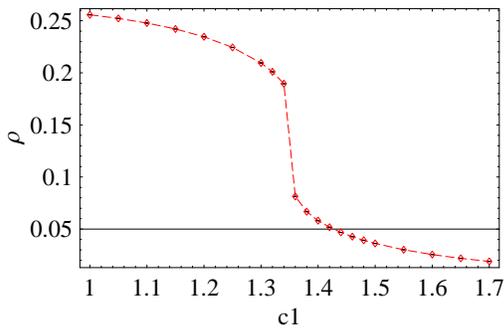}
\end{center}
\caption{Instanton density for $c_2=0.5$ as a function of $c_1$.
At the phase transition point $c_1\simeq 1.35$, it changes its behavior.
System size $L=24$.}
\label{Instanton-z2c205}
\end{figure}
\begin{figure}[bh]
\begin{center}
\includegraphics[width=7cm]{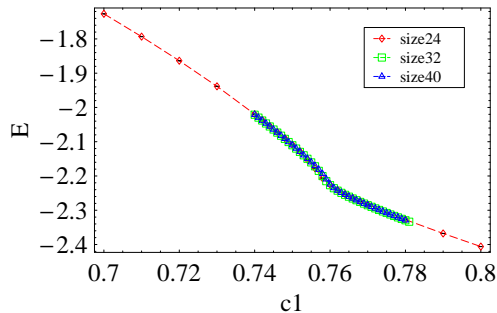}
\end{center}
\caption{$E$ for $c_1=0.3$ as a function of $c_2$.
System size $L=24,\;32,\;40$.
There is almost no system-size dependence.}
\label{E-z2c103}
\end{figure}

We also measured the instanton density $\rho(x)$ for $c_2=0.5$ as a function 
of $c_1$.
$\rho(x)$ is defined as follows for the gauge field configuration 
$U_{x,\mu}=e^{i\theta_{x,\mu}},\; \theta_{x,\mu}=0,\; \pi$\cite{Inst-1,CPN-2}.
First we consider the magnetic flux $\Theta_{x,\mu\nu}$ penetrating plaquette
$(x,x+\mu,x+\mu+\nu,x+\nu)$
\begin{eqnarray}
\Theta_{x,\mu\nu}&=&\theta_{x,\mu}+\theta_{x+\mu,\nu}-\theta_{x+\nu,\mu}
-\theta_{x,\nu}, \nonumber \\
 && (-2\pi\le \Theta_{x,\mu\nu}\le 2\pi).
\label{Flux}
\end{eqnarray}
We decompose $\Theta_{x,\mu\nu}$ into its integer part $n_{x,\mu\nu}$,
which represents the Dirac string (vortex line), and the remaining part 
$\tilde{\Theta}_{x,\mu\nu}$,
\begin{equation}
\Theta_{x,\mu\nu}=2\pi n_{x,\mu\nu}+\tilde{\Theta}_{x,\mu\nu}, \;\;
(-\pi\le \tilde{\Theta}_{x,\mu\nu}\le \pi).
\label{Flux2}
\end{equation}
Then instanton density $\rho(x)$ at the cube around the site
$x+{\hat{0} \over 2}+{\hat{1} \over 2}+{\hat{2} \over 2}$ of the
dual lattice is defined as 
\begin{equation}
\begin{split}
\rho(x) &=-{1 \over 2}\sum_{\mu\nu\lambda}\epsilon_{\mu\nu\lambda}
(n_{x+\mu,\nu\lambda}-n_{x,\nu\lambda})  \\
&={1 \over 4\pi}\sum_{\mu\nu\lambda}\epsilon_{\mu\nu\lambda}
(\tilde{\Theta}_{x+\mu,\nu\lambda}-\tilde{\Theta}_{x,\nu\lambda}),
\end{split}
\label{rho}
\end{equation}
where $\epsilon_{\mu\nu\lambda}$ is the antisymmetric tensor.
From the above definition, it is obvious that 
$\rho\equiv\langle |\rho(x)| \rangle$ measures probability of
creation/annihilation of magnetic vortex.
In $3D$ $Z_2$ gauge theory, magnetic vortices in $3D$ can be regarded as 
world lines of flux quanta dubbed vison.
Nonvanishing value of $\rho$ means that the number of visons is not conserved,
and therefore {\em condensation of the vison}.
The result of calculation of $\rho$ is shown in Fig.\ref{Instanton-z2c205}.
There is a sharp discontinuity at the phase transition $c_1\simeq 1.35$.
In phase I, finite value of $\rho$ means large fluctuations of the
gauge field and spinon $v_x$ is confined to gauge-invariant composites,
$\vec{n}_i\; (i=1,2,3)$.
This phenomenon is sometimes called {\em dual Meissner effect}.
On the other hand in phase II, $\rho$ is strongly suppressed and
the topological order exists.
Later study on the spin correlation function reveals that 
this suppression is due to Higgs mechanism by the condensation of $v_x$.

\begin{figure}[h]
\begin{center}
\includegraphics[width=7cm]{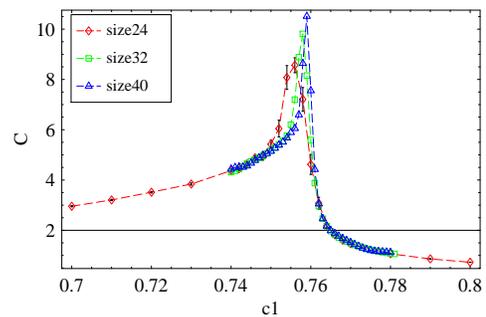}
\end{center}
\caption{$C$ for $c_1=0.3$ as a function of $c_2$ with $L=24,\;32,\;40$.
Its system-size dependence indicates that the phase transition is
of second order. In addition to the standard MC simulations, we used
multi-histogram methods to obtain reliable values of $C$ near the 
phase transition point\cite{MHM}.}
\label{C-z2c103}
\end{figure}
\begin{figure}[h]
\begin{center}
\includegraphics[width=7cm]{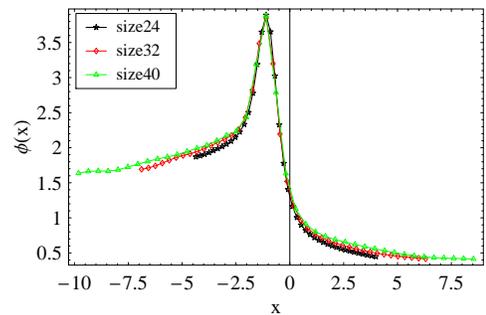}
\end{center}
\caption{FSS for $c_1=0.3$. All data for $L=24,\; 32,\; 40$ can be
fit by single function $\phi(x)$.}
\label{FSS-z2c103}
\end{figure}

Next we consider the phase transition from phase I to III.
We show $E$ and $C$ for $c_1=0.3$ in Figs.\ref{E-z2c103} and
\ref{C-z2c103}.
The results indicate that there exists a second-order phase transition at
$c_2\simeq 0.76$.
By the finite-size scaling (FSS) hypothesis for $C$,
\begin{equation}
C_L(\epsilon)=L^{\sigma/\nu}\phi(L^{1/\nu}\epsilon),
\label{FSS}
\end{equation}
where $C_L$ is the ``specific heat" of system size $L$, and
$\epsilon\equiv (c_2-c_{2\infty})/c_{2\infty}$ with $c_{2\infty}$
(the critical coupling for $L\rightarrow \infty$),
we estimated the critical exponents $\nu,\; \sigma$ by using 
the FSS (\ref{FSS}) and obtained $\nu=0.63,\; \sigma=0.17$ and the critical
coupling $c_{2\infty}=0.76$.
The obtained scaling function $\phi(x)$ is shown in Fig.\ref{FSS-z2c103}.
These values are very close to those of the pure $Z_2$ gauge model
that are obtained from the data of the $3D$ Ising model by duality\cite{dual}.

\begin{figure}[th]
\begin{center}
\includegraphics[width=7cm]{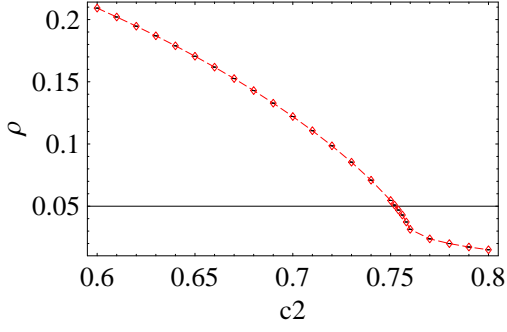}
\end{center}
\caption{Instanton density for $c_1=0.3$ as a function of $c_2$.
At the phase transition point $c_2\simeq 0.76$, it changes its behavior.
System size $L=24$.}
\label{Instanton-z2c103}
\end{figure}

We also measured instanton density $\rho$ and show the result in 
Fig.\ref{Instanton-z2c103}.
$\rho$ is a decreasing function of $c_2$ and changes its behavior
at the phase transition point $c_2\simeq 0.76$.

\begin{figure}[bh]
\begin{center}
\includegraphics[width=7cm]{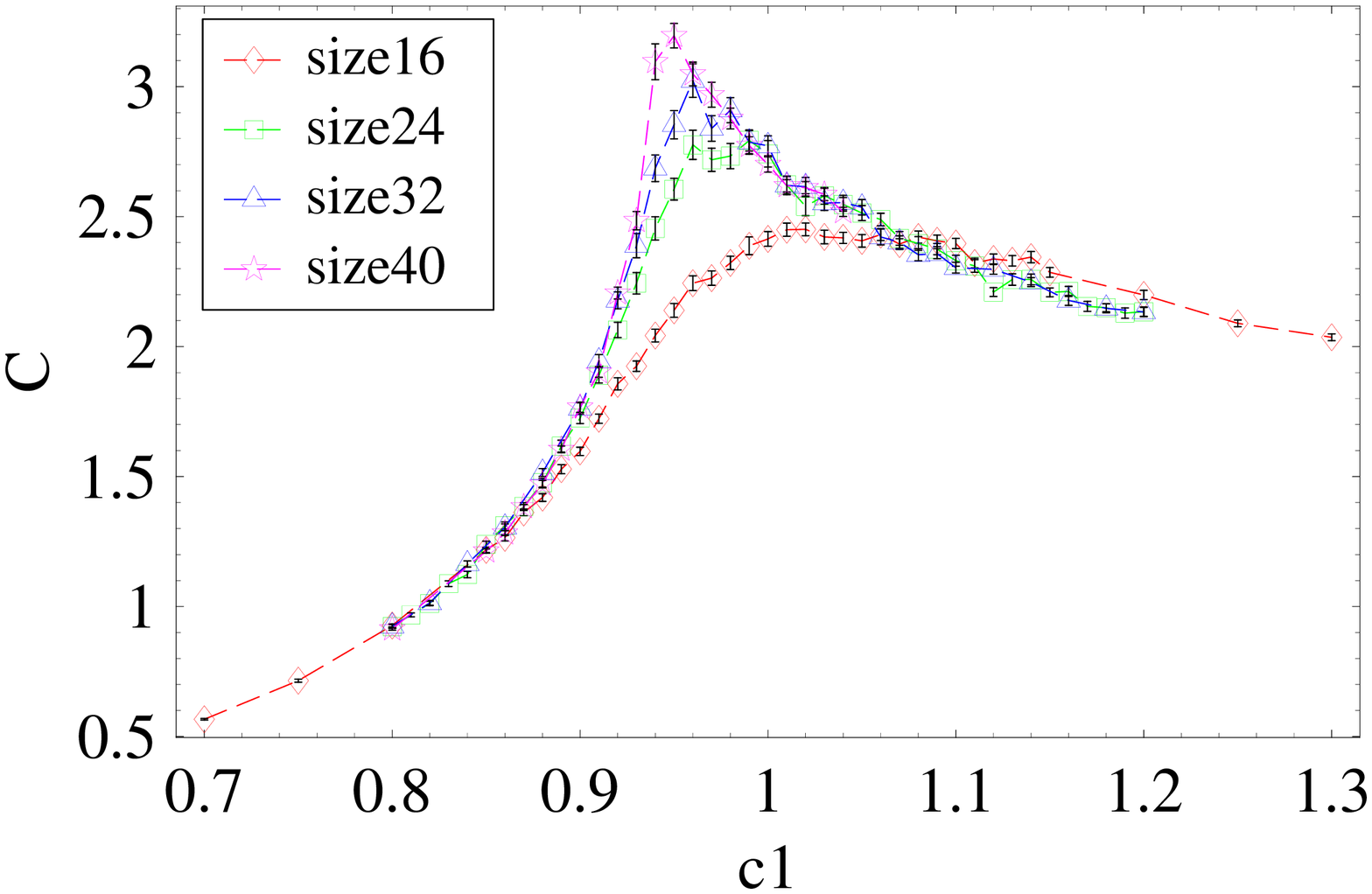}
\end{center}
\caption{$C$ for $c_2=1.5$ as a function of $c_1$ with $L=16,\;24,
\;32, \;40$.
Its system-size dependence indicates that the phase transition is
of second order.}
\label{C-z2c215}
\end{figure}

Finally, let us consider the phase transition from the phases II to III.
Obtained $E$ has no system-size dependence.
System-size dependence of $C$ is shown in Fig.\ref{C-z2c215}, from which
we judge that the phase transition is of second order.
By the FSS, the critical exponents are estimated as $\nu=0.65,\;
\sigma=0.156,\; c_{2\infty}=0.93$.
This value of $\nu$ should be compared with that of the $O(4)$
nonlinear sigma model in $3D$, $\nu_{O(4)}=0.75$.
At present, it is not clear for us if the above two phase transitions belong to
the same universality class.

\begin{figure}[th]
\begin{center}
\includegraphics[width=7cm]{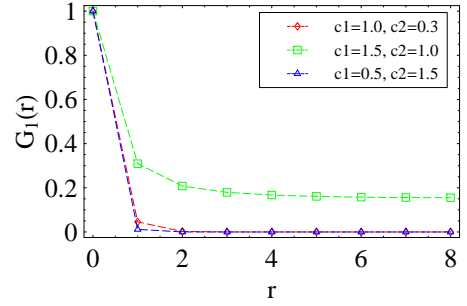}
\end{center}
\caption{Correlation function of the spin field $\vec{n}_1(x)$.
It has a long-range order only in phase II.}
\label{Scf-n1-s16}
\end{figure}

\begin{figure}[th]
\begin{center}
\includegraphics[width=7cm]{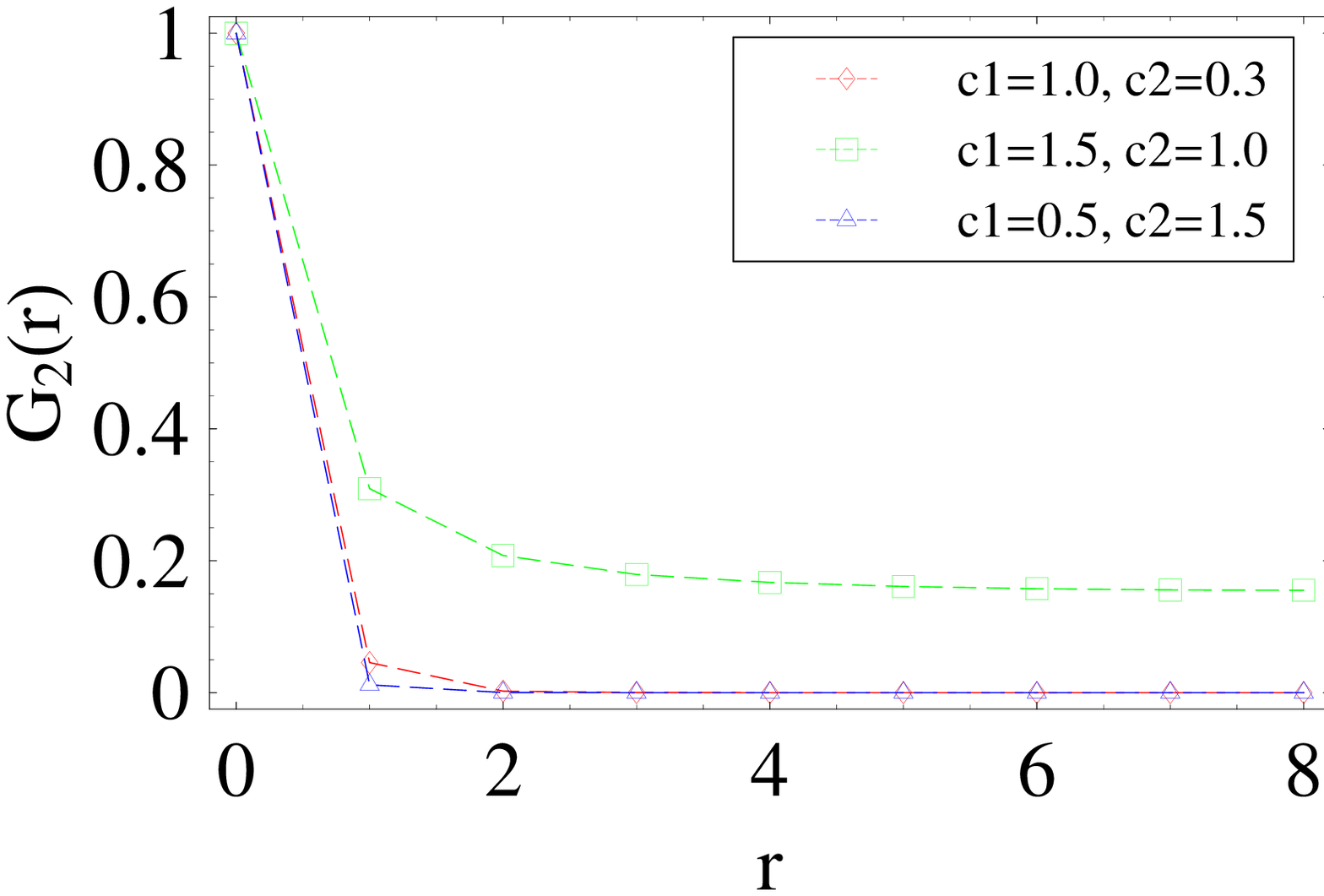}
\end{center}
\caption{Correlation function of the spin field $\vec{n}_2(x)$.
It has a long-range order only in phase II.}
\label{Scf-n2-s16}
\end{figure}

\begin{figure}[bh]
\begin{center}
\includegraphics[width=7cm]{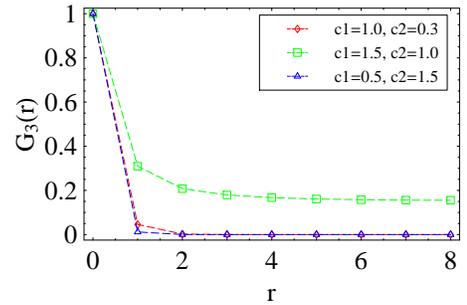}
\end{center}
\caption{Correlation function of the spin-nematic field $\vec{n}_3(x)$.
It has a long-range order only in phase II.}
\label{Scf-n3-s16}
\end{figure}

In order to see (non)existence of the magnetic long-range order (LRO), 
we measured correlation functions of the spins 
$\vec{n}_1(x),\; \vec{n}_2(x)$ and $\vec{n}_3(x)$.
They are defined as, 
\begin{equation}
G_i(r)={1 \over L^3}\sum_x 
\langle \vec{n}_i(x+r)\cdot \vec{n}_i(x)\rangle, \;\;\; i=1,2,3.
\label{SCF}
\end{equation}
We exhibit the results in Figs.\ref{Scf-n1-s16}, \ref{Scf-n2-s16}
and \ref{Scf-n3-s16}.
It is obvious that only in phase II, they have the LRO.
This LRO indicates a nonvanishing expectation value of $v_x$,
$\langle v_x \rangle \neq 0$, in phase II.
This understanding is supported by the measure of $\rho$, which 
shows that the gauge dynamics is in the Higgs phase in phase II.

\begin{figure}[h]
\begin{center}
\includegraphics[width=7cm]{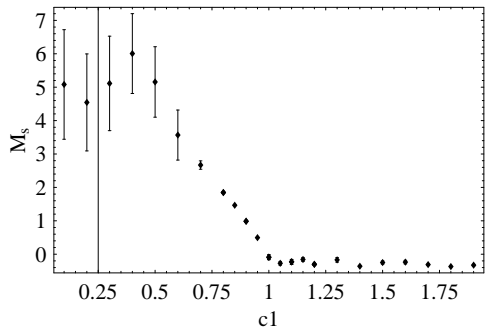}
\end{center}
\caption{Spin gap as a function of $c_1$ for $c_2=1.5$.}
\label{Seg}
\end{figure}

We also calculated the spin gap as a function $c_1$ for $c_2=1.5$,
i.e., from the spin liquid to spiral state.
It is difficult to estimate the spin gap directly from the spin
correlation functions $G_i(r)$. 
Then as in the previous studies\cite{CPN-2}, 
we employ a Fourier transformation
of the spin field, e.g., $\vec{n}_3(x)$,
\begin{equation}
\tilde{\vec{n}}_3(x_0;p_1,p_2)=
\sum_{x_1,x_2}e^{ip_1x_1+ip_2x_2}\vec{n}_3(x).
\label{Fourier}
\end{equation}
In the continuum limit, the correlator of $\tilde{\vec{n}}_3(x_0;p_1,p_2)$
behaves as 
\begin{eqnarray}
&& \langle \tilde{\vec{n}}_3(x_0;p_1,p_2)\cdot \tilde{\vec{n}}_3(0;p_1,p_2)
\rangle \nonumber  \\
&& \hspace{0.5cm} =\int \; dp_0{e^{ip_0x_0} \over \vec{p}^2+M^2_s} 
        \propto e^{-\sqrt{p^2_1+p^2_2+M^2_s}x_0},
\label{Ms}
\end{eqnarray}
where $\vec{p}^2=\sum_{i=1,2,3}p^2_i$.
In the practical calculation on the lattice, we put $p_1=p_2=2\pi/L$,
and measured $\sqrt{p^2_1+p^2_2+M^2_s}$ from the correlation function
of $\tilde{\vec{n}}_3(x_0;p_1,p_2)$.
We show the result $M_s$ as a function of $c_1$ in Fig.\ref{Seg}.
It is obvious that the spin gap $M_s$ is a continuous decreasing function
of $c_1$ and is vanishing for $c_1>c_{1c}\simeq 1$.
This result means that the spin excitation has a finite gap in the $Z_2$
spin liquid, whereas the spin wave in the spiral state is gapless.

From all the above calculations, we obtain the phase diagram shown in 
Fig.\ref{phase-z2}.


\subsection{U(1) gauge field coupled to $Z_2$CP$^1$ spinon:
$c_3=0$ case}
Let us consider the case $c_3=0$ of the system (\ref{Z2action1}).
The system with $c_1=0$ is nothing but the pure compact U(1) gauge
model in $3D$.
It is well-known that there is no phase transition and the system is
always in the confined phase, though there is a crossover from
dense-instanton to dilute-instanton regimes as the parameter $c_2$
is increased.

\begin{figure}[th]
\begin{center}
\includegraphics[width=7cm]{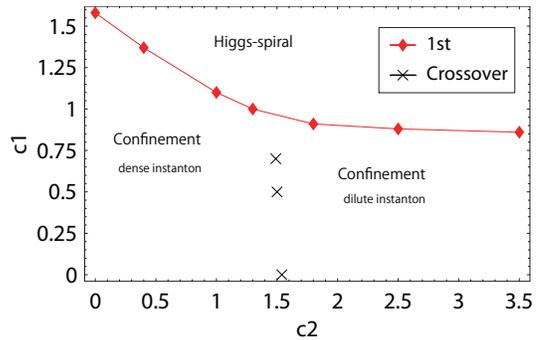}
\end{center}
\caption{Phase diagram of the U(1) gauge theory of CP$^1$ spinons
with $c_3=0$.
There are two phases.}
\label{phase-u1}
\end{figure}
\begin{figure}[h]
\begin{center}
\includegraphics[width=7cm]{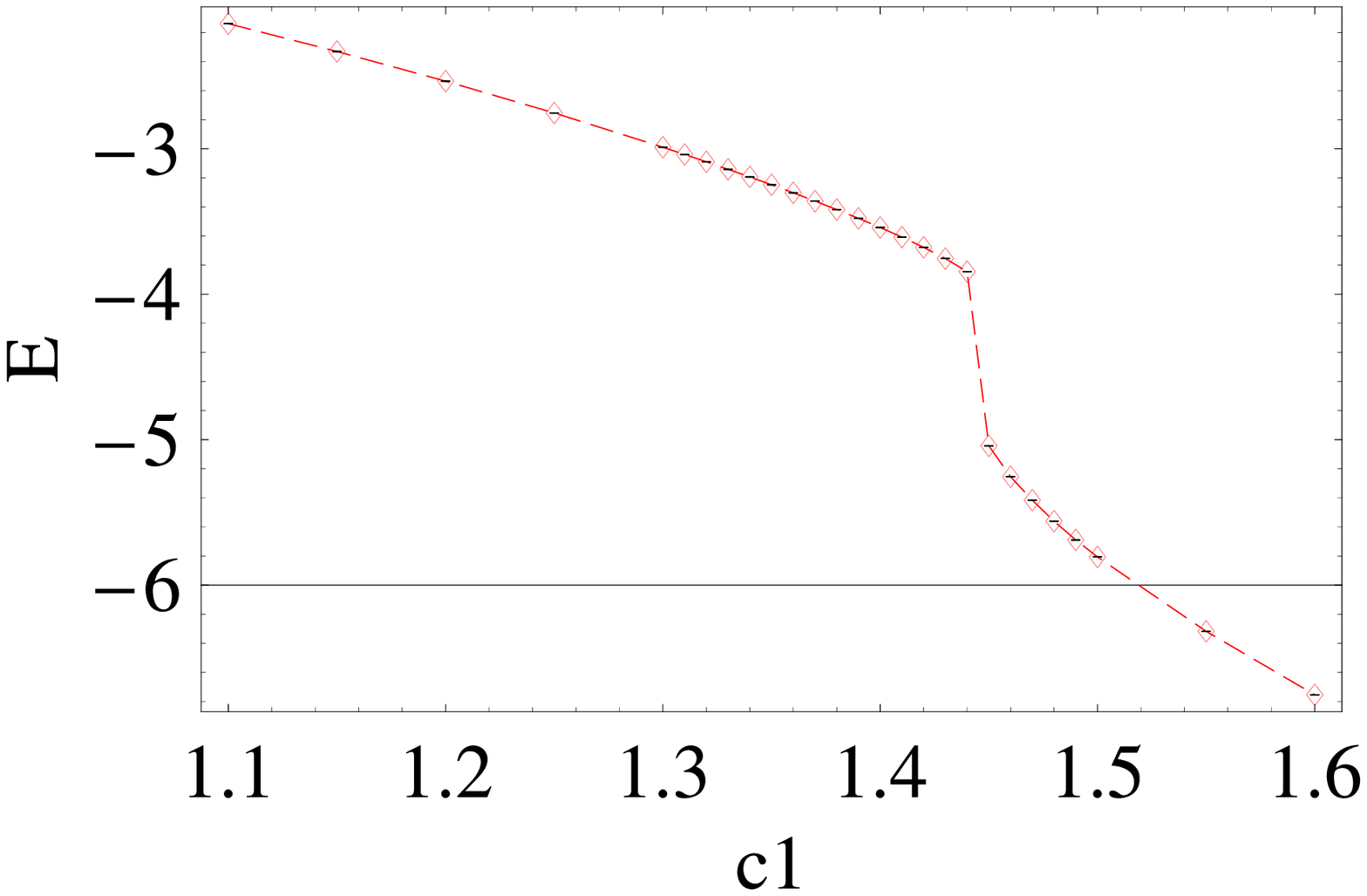}
\end{center}
\caption{$E$ for $c_2=0.4$. There is sharp discontinuity at $c_1=1.44$
that indicates a first-order phase transition. System size is $L=24$.}
\label{E-u1c204c300}
\end{figure}
\begin{figure}[bh]
\begin{center}
\includegraphics[width=7cm]{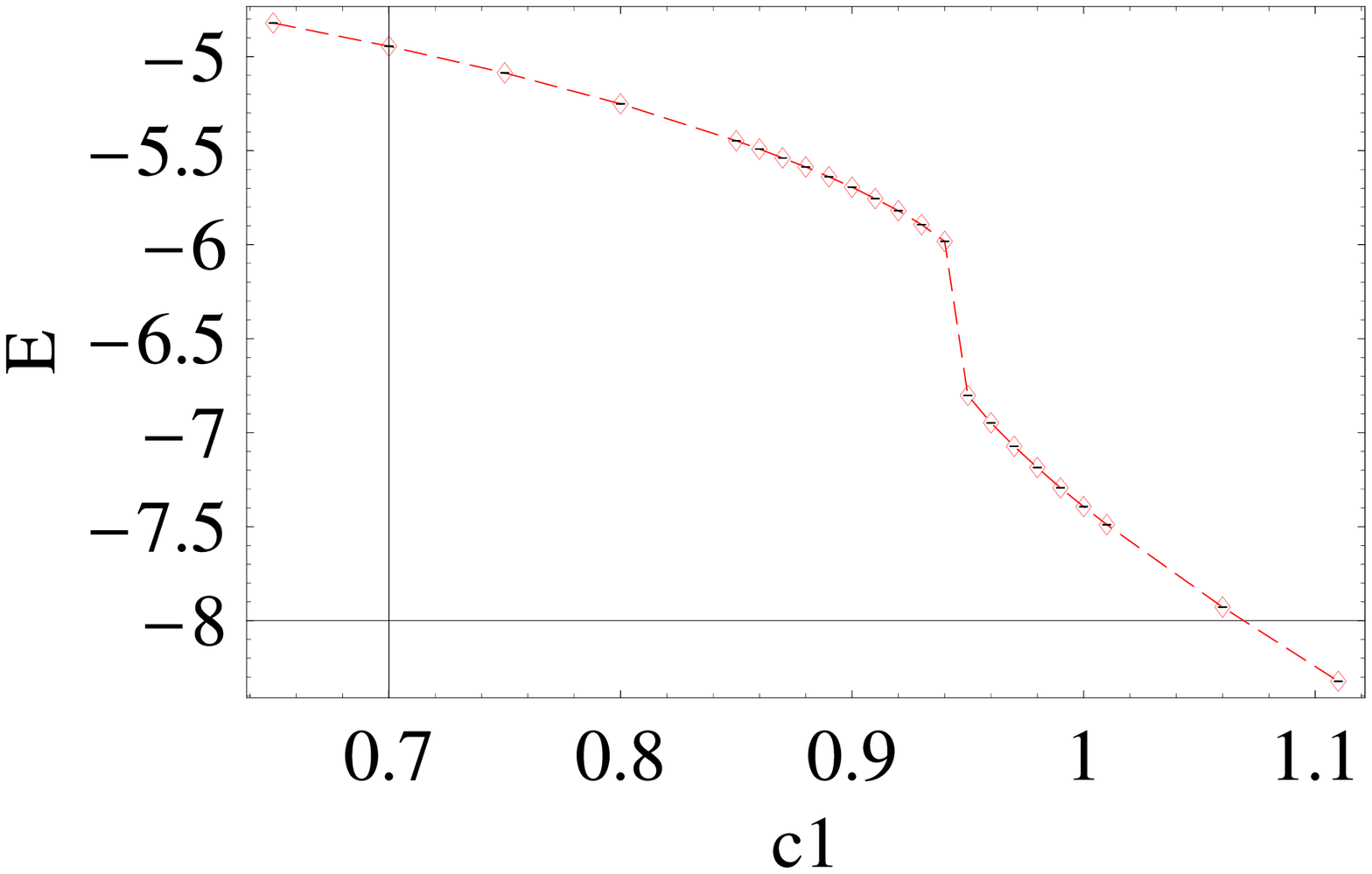}
\end{center}
\caption{$E$ for $c_2=1.8$. There is sharp discontinuity at $c_1=0.95$
that indicates a first-order phase transition. System size is $L=24$.}
\label{E-u1c218c300}
\end{figure}
\begin{figure}[h]
\begin{center}
\includegraphics[width=5cm]{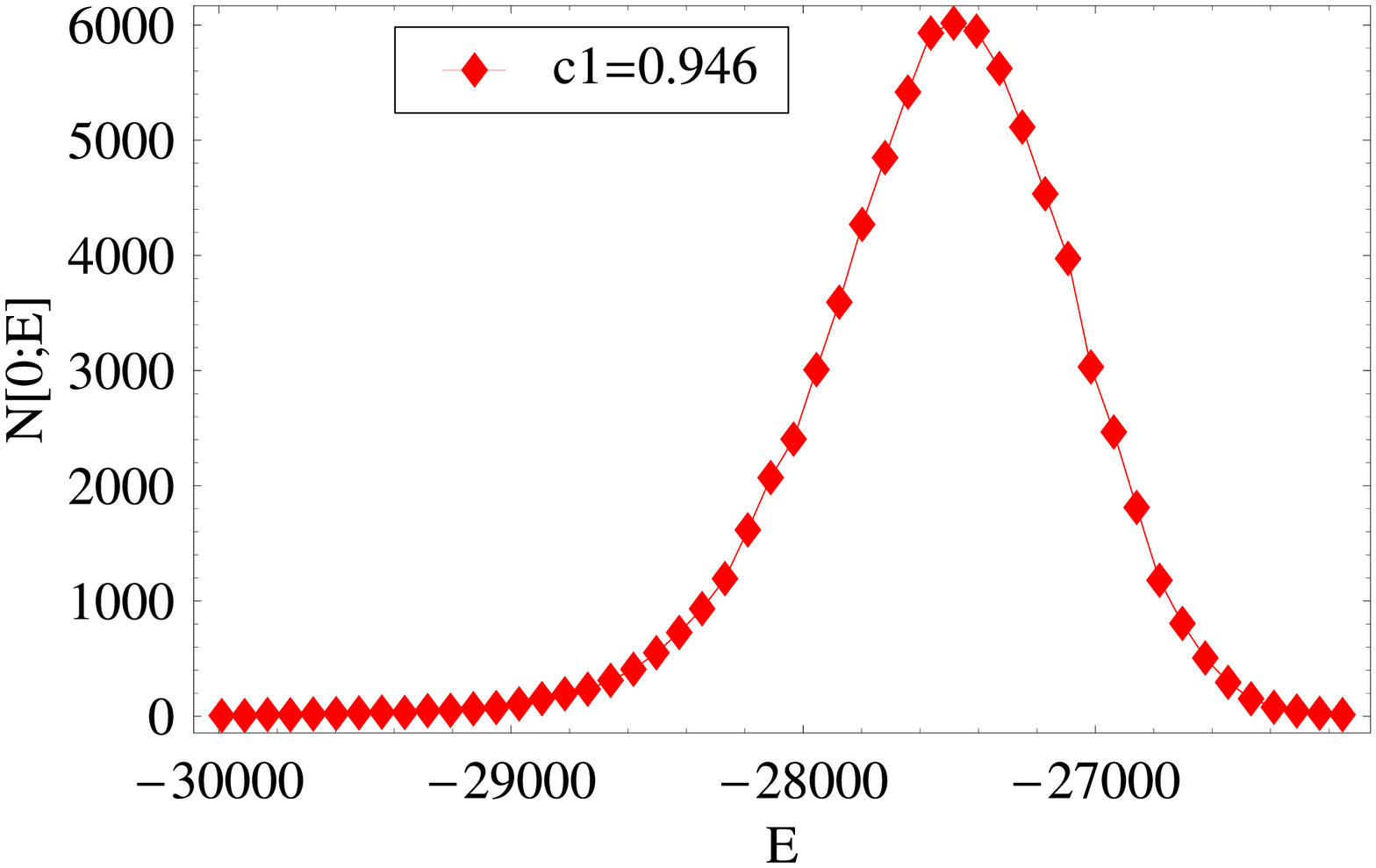}
\includegraphics[width=5cm]{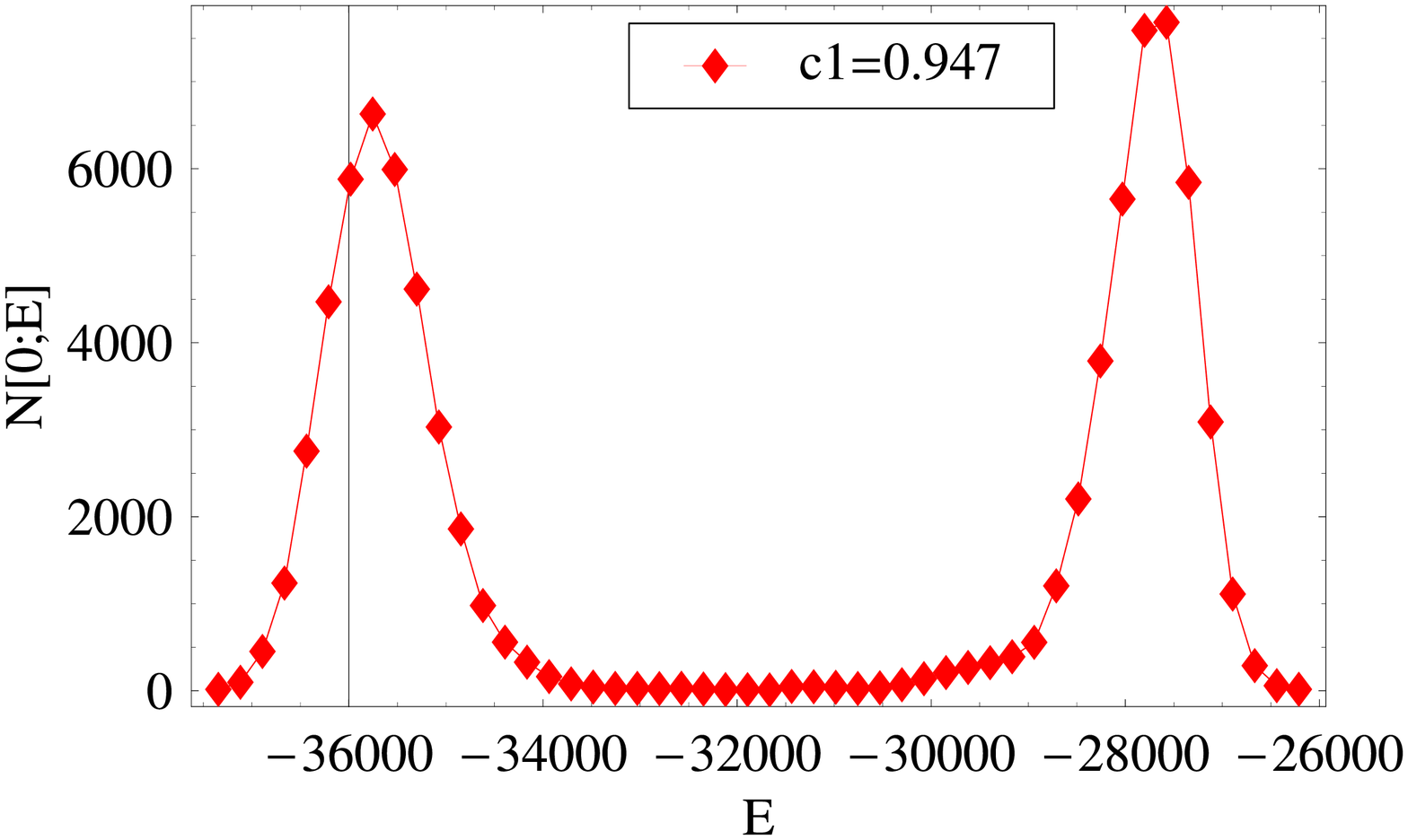}
\includegraphics[width=5cm]{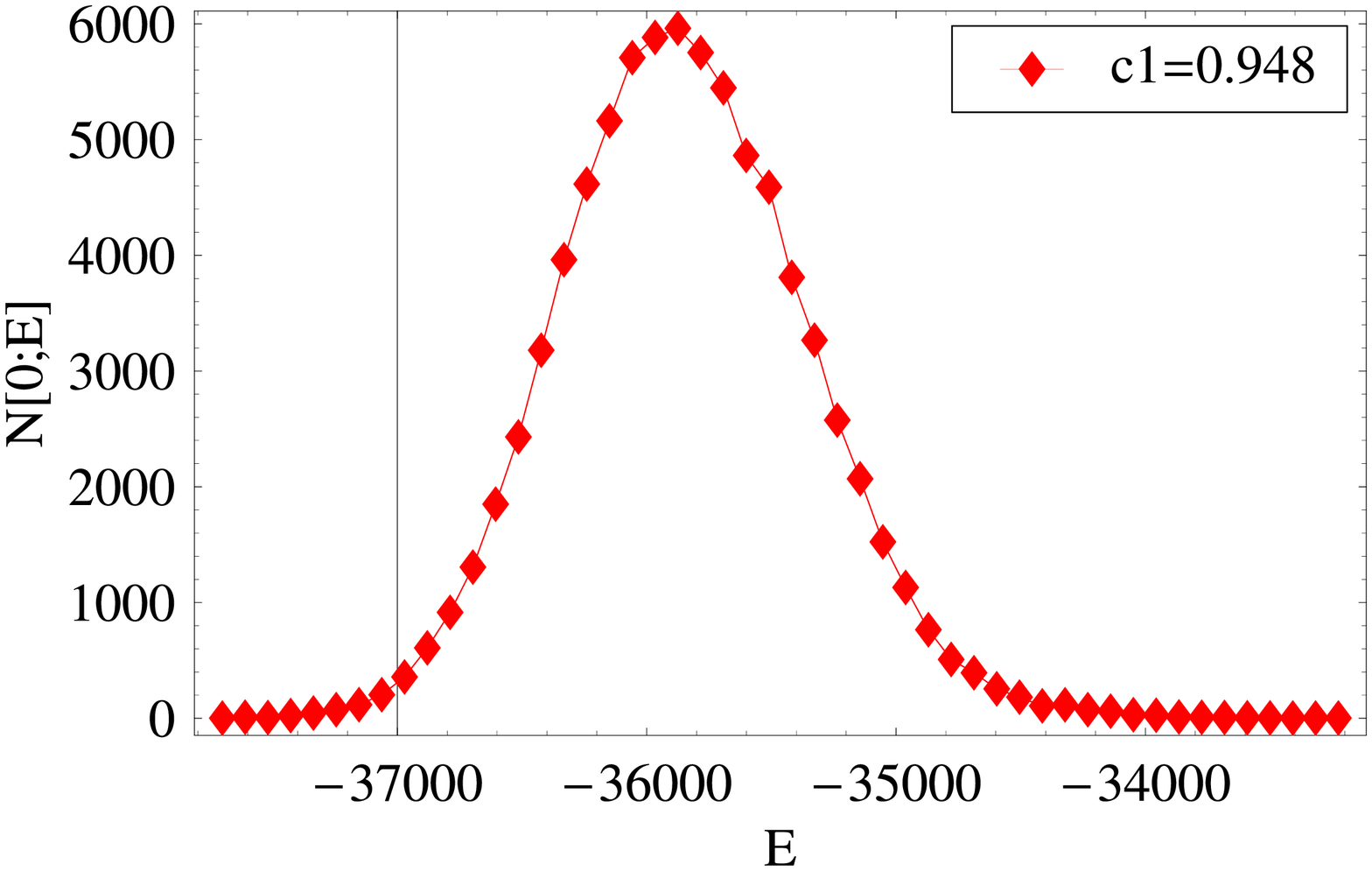}
\end{center}
\caption{Distribution of $A(0)$ for $c_2=1.8$ and $c_1$ close to
the phase transition point. The double-peak structure at $c_1=0.947$ 
confirms the existence of the first-order phase transition.
System size $L=24$.}
\label{hist-u1c300}
\end{figure}
\begin{figure}[h]
\begin{center}
\includegraphics[width=7cm]{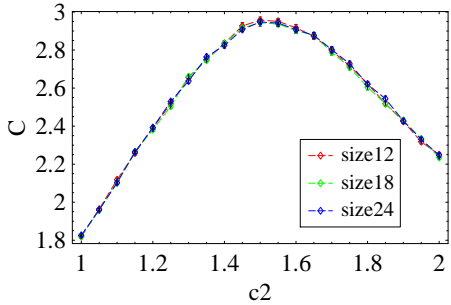}
\end{center}
\caption{$C$ for $c_1=0.5$. System size $L=12,\;18,\;24$.
There is no system-size dependence, i.e., the second-order
phase transition in the $Z_2$ gauge model reduces to a crossover.
The deconfined spin-liquid phase does not exists in the model with
$c_3=0$.}
\label{C-u1c105}
\end{figure}
\begin{figure}[th]
\begin{center}
\includegraphics[width=7cm]{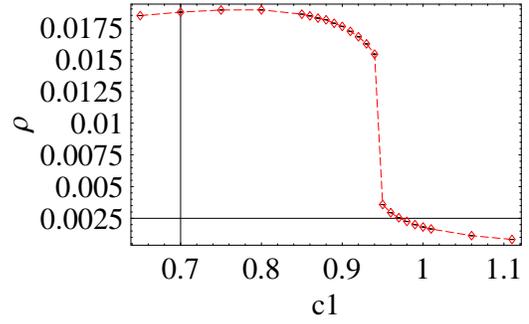}
\end{center}
\caption{Instanton density for $c_2=1.8$ as a function of $c_1$.
At the phase transition point $c_1\simeq 0.95$, there is a sharp
discontinuity. System size $L=24$.}
\label{Instanton-u1c218c300}
\end{figure}
\begin{figure}[h]
\begin{center}
\includegraphics[width=6cm]{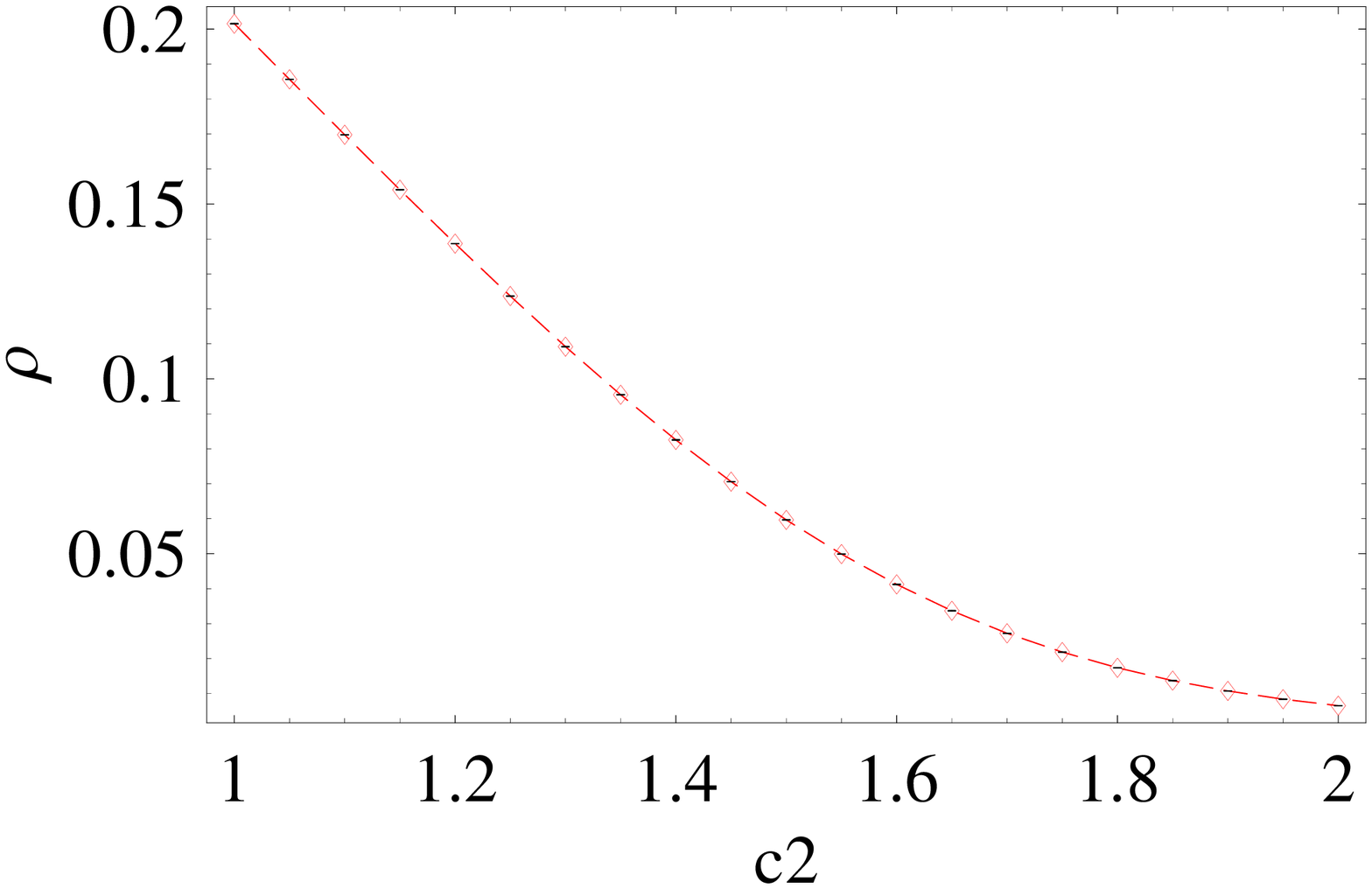}
\end{center}
\caption{Instanton density for $c_1=0.5$ as a function of $c_2$.
There is no anomalous behavior at the crossover $c_2\simeq 1.5$.
System size $L=24$.}
\label{Instanton-u1c105c300}
\end{figure}

We show the obtained phase diagram of the system (\ref{Z2action1}) 
with $c_3=0$ in Fig.\ref{phase-u1}.
There are two phases, (i) phase I is the dimer phase with confinement of 
spinons and without any long-range order, (ii) phase II is the spiral
state with the condensation of $v_x$.
Phase transitions separating these two phases are of first order
as the calculations of $E$ in Figs.\ref{E-u1c204c300} and
\ref{E-u1c218c300} indicates.
In order to verify this observation, we measured
the distribution of $A(0)$, ${\cal N}[0;E]$, in the MC steps.
From Fig.\ref{hist-u1c300}, it is obvious that on the critical line 
${\cal N}[0;E]$
has a double-peak structure whereas it does not off the critical line.

The second-order phase transition that exists in the $Z_2$ gauge model of
spinons from the confined to deconfined phases disappears in the 
system with $c_3=0$.
There is a crossover line emanating from the crossover point of the
pure compact U(1) gauge model in $3D$.
See Fig.\ref{C-u1c105}.
This fact strongly influences structure of the ground state and 
low-energy excitations of the original spin system.
Absence of the deconfined phase means that {\em the spin liquid phase
does not exist} in the present case.
We measured the instanton density$\rho$ to verify the above conclusion.
In particular in the dilute-instanton regime $c_2>1.5$, the
instanton density $\rho$ is small even in the confined phase
$c_1<c_{1c}(c_2)$, but it decreases rapidly at the phase transition
$c_1=c_{1c}(c_2)$ to the Higgs phase.
See Fig.\ref{Instanton-u1c218c300}.
On the other hand for $c_1=0.5$, $\rho$ is a decreasing function of
$c_2$ but does not exhibit any anomalous behavior at the crossover 
$c_2\simeq 1.5$.
See Fig.\ref{Instanton-u1c105c300}.

We calculated the spin correlation functions in each phase 
and verified that the spin LRO exists only in the Higgs phase 
$c_1>c_{1c}$.

\subsection{Massive U(1) gauge field coupled with $Z_2$CP$^1$ spinon:
$c_3=1.0$ case}

\begin{figure}[h]
\begin{center}
\includegraphics[width=7cm]{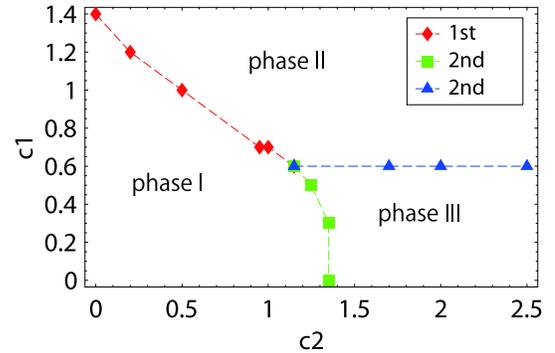}
\end{center}
\caption{Phase diagram of the U(1) gauge theory of CP$^1$ spinons
with $c_3=1.0$. There are three phases as in the $Z_2$ gauge model.
}
\label{phase-05}
\end{figure}
\begin{figure}[h]
\begin{center}
\includegraphics[width=7cm]{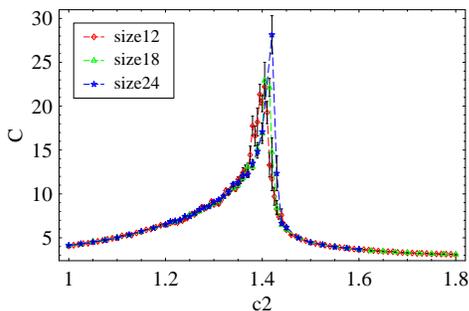}
\end{center}
\caption{$C$ for $c_1=0.3$. System size $L=12,\;18,\;24$.
}
\label{C-u1c103}
\end{figure}
\begin{figure}[bh]
\begin{center}
\includegraphics[width=7cm]{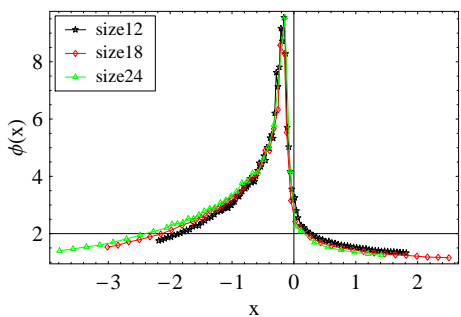}
\end{center}
\caption{FSS for $c_1=0.3$. All data for $L=12,\; 18,\; 24$ can be
fit by single function $\phi(x)$.}
\label{FSS-u1c103}
\end{figure}
\begin{figure}[t]
\begin{center}
\includegraphics[width=7cm]{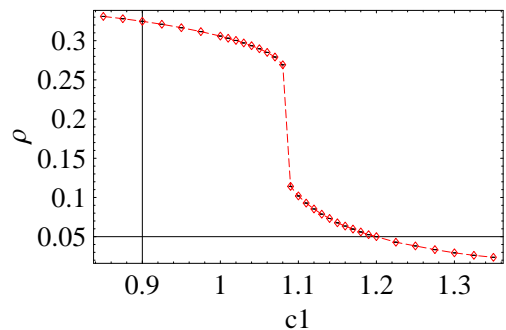}
\end{center}
\caption{Instanton density for $c_2=0.5$ as a function of $c_1$.
At $c_1 \simeq 1.1$, there is sharp discontinuity corresponding to
the first-order phase transition.
System size $L=24$.}
\label{Instanton-u1c205c305}
\end{figure}
\begin{figure}[t]
\begin{center}
\includegraphics[width=7cm]{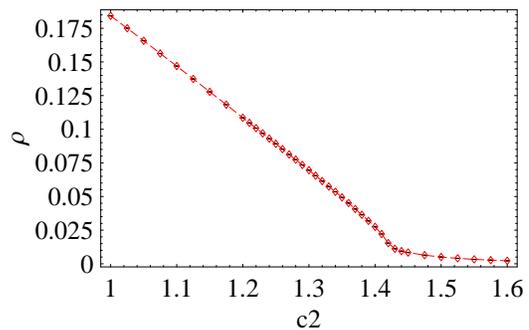}
\end{center}
\caption{Instanton density for $c_1=0.3$ as a function of $c_2$.
At $c_2\simeq 1.4$, $\rho$ changes its behavior.
System size $L=24$.}
\label{Instanton-u1c103c305}
\end{figure}

Finally let us consider the case $c_3=1.0$.
We also investigated the case $c_3=2.0$ and obtained similar results.
In the present case $c_3>0$, the gauge field $U_{x,\mu}$ is a $U(1)$ variable, 
but local U(1) gauge symmetry is explicitly broken down to $Z_2$ by both 
the hopping term of $v_x$ and the mass term of the gauge field, i.e.,
$U_{x,\mu}^2+\mbox{c.c.}$
It is expected that the mass term of the gauge field is a relevant
perturbation and therefore the phase structure of the system $c_3>0$
is qualitatively the same with that of the $Z_2$ gauge theory studied in
Sec.III.A.

We show the obtained phase diagram in Fig.\ref{phase-05}.
There are three phases similarly to the $Z_2$ gauge theory of spinons,
as it is expected.
In Figs.\ref{C-u1c103} and \ref{FSS-u1c103}, 
we show $C$ as a function of $c_2$ for $c_1=0.3$
and the FSS scaling function obtained from these data.
Critical exponents are estimated as $\nu=1.26,\; \sigma=0.43$ and 
$c_{2\infty}=1.44$.
From the above result, we think that the present phase transition
does not belong to the universality class of the $3D$ Ising model.
At phase transitions from the tilted-dimer to spiral phases,
$E$, the distribution of ${\cal N}[1;E]$ and the instanton density
$\rho$ have similar behaviors to those in  the previous cases of 
the first-order phase transition.
In Figs.\ref{Instanton-u1c205c305} and \ref{Instanton-u1c103c305},
we show the result of the instanton density.


\section{Conclusion and discussion}

In the present paper, we derived effective gauge models 
that describe low-energy properties of antiferromagnets with
frustrations in $2D$, and studied their phase structure mostly by means of 
MC simulations.
We found that generally there are three phases in the models, 
(i) phase of the tilted dimer state with spin-triplet excitations, 
(ii) the spiral state with gapless spin wave,
(iii) the spin liquid with weakly interacting spinons.
We identified the order of the phase transitions and estimated
values of the critical exponents of the second-order phase transitions.
The investigation suggests that for the spin liquid to appear, 
multi-spin and nonlocal interactions are necessary in the original
spin systems.

In order to verify the validity of the above results,
it is important and also possible to study spin systems on
layered $3D$ triangular lattice at finite temperature ($T$) by means of the 
Schwinger boson (CP$^1$) representations. 
In this case, the systems can be studied directly with the spatial
lattice as a regularization.
In the path-integral representation of the partition function 
$Z$ in (\ref{Z}) at finite $T$, the $\tau$-dependence of $z_i$ is ignored.
Then the path-integral over CP$^1$ variables $z_i$'s in $Z$ can be performed
without any difficulties by the MC simulations.
At present we are studying these systems, and have obtained preliminary
results that support the conclusion in the present paper\cite{NKIM}.

In the present paper, we mostly focus on the (short-range) spiral
state with $\langle \Lambda_{ij}\rangle \propto 
\langle z_i\cdot \bar{\tilde{z}}_j\rangle=\Lambda_0\neq 0$.
There is another possibility of canted state like 
$\langle \Lambda_{i,i+\hat{1}} \rangle =(-)^i\Lambda_0\neq 0$.
This state can be regarded as a state with a ferromagnetic order in the 
AF background.
This state also breaks the U(1) gauge invariance down to the $Z_2$ as the
(short-range) spiral state does, and therefore results obtained 
in the present paper are expected to be applicable to the canted state.

Finally let us comment on effects of the Berry phase.
As we explained in Sec.II, the Berry phase appears after 
integrating out the high-energy modes in the path integral in order
to derive the effective field theory of AF magnets.
The Berry phase may play an important role though qualitative phase 
structure is not changed by its existence\cite{FNBerry1}.
Whether the suppression of the instantons occurs by the Berry phase
strongly depends on its coefficient.
For example in the {\em inhomogeneous} AF Heisenberg model on a square lattice,
the coefficient depends on the magnitude of the inhomogeneity
and is generally an irrational\cite{Yoshi}.
Suppression of instantons does not occur in that case and 
the N\'eel-dimer phase
transition belongs to the universality class of the classical 
$3D\; O(3)$ nonlinear
sigma model, which is equivalent to the CP$^1$ gauge model (\ref{Sz})
{\em without} the Berry phase.
This result was verified by the numerical study of the inhomogeneous
SU(2) AF Heisenberg model.
We expect that nonvanishing frustration coupling $J'$ gives a similar
effect on the Berry phase's coefficient because the most dominant
NN spin pair configuration is shifted from 
$z_{ia}=\epsilon_{ab}\bar{z}_{jb}$ ($i,j=$ site, $a,b=$ spinor indices)
on path-integrating out high-energy modes.
If this is the case, the Berry phase gives only negligible effects
on critical behavior of the systems under study, and the FSS used
in the present paper gives reliable estimation of the critical 
exponents\cite{FNBerry2}.

\bigskip
\acknowledgments
This work was partially supported by Grant-in-Aid
for Scientific Research from Japan Society for the 
Promotion of Science under Grant No.20540264.



\begin{references} 

\bibitem{LNW}See for example, P.A.Lee, N.Nagaosa, and X.-G.Wen, \\
Rev.Mod.Phys.{\bf 78}, 17(2006).

\bibitem{AF1}K.Miyagawa, A.Kawamoto, Y.Nakazawa, and \\
K.Kanoda, Phys. Rev. Lett.{\bf 75}, 1174(1995).

\bibitem{AF2-1}T.Nakamura, T.Takahashi, S.Aonuma, and R.Kato,\\
J. Mater. Chem.{\bf 11}, 2159(2001).

\bibitem{AF2-2}
M.Tamura, and R.Kato, \\
J. Phys.: Condens. Matter {\bf 14}, L729(2002).

\bibitem{AF2-3}
Y.Shimizu, H.Akimoto, H.Tsuji, A.Tajima, and R.Kato,
Phys. Rev. Lett.{\bf 99}, 256403(2007).

\bibitem{AF3-1}Y.Shimizu, K.Miyagawa, K.Kanoda, M.Maesato, and \\
 G.Saito, Phys. Rev. Lett.{\bf 91}, 107001(2003).

\bibitem{AF3-2}
S.Yamashita, Y.Nakazawa, M.Oguni, Y.Oshima, \\
H.Nojiri, Y.Shimizu, K.Miyagawa, and K.Kanoda, \\
Nature Physics {\bf 4}, 459(2008).

\bibitem{Qi}Y.Qi, C.Xu, and S.Sachdev, \\
Phys. Rev. Lett.{\bf 102}, 176401(2009).

\bibitem{CCC1}R.Coldea, D.A.Tennant, A.M.Tsvelik, and Z.Tylczynski,\\
Phys.Rev.Lett.{\bf 86}, 1335(2001).

\bibitem{CCC2}R.Coldea, D.A.Tennant, and Z.Tylczynski,\\
Phy.Rev.B {\bf 68}, 134424(2003).

\bibitem{Sachdev1}See for example, S.Sachdev,
Nature Physics {\bf 4}, 173(2008), and references cited therein.

\bibitem{CP-1}D.Arovas and A.Auerbach,
Phys. Rev.B {\bf 38}, 316(1988).

\bibitem{CP-2}I.Ichinose and T.Matsui,
Phys. Rev.B {\bf 45}, 9976(1992).

\bibitem{Sawa}K.Sawamura, T.Hiramatsu, K.Ozaki, I.Ichinose, and \\
T.Matsui, Phys.Rev.B {\bf 77}, 224404(2008).

\bibitem{Berry-1}F.D.M.Haldane,
Phy. Rev. Lett.{\bf 61}, 1029(1988).

\bibitem{Berry-2}N.Read and S.Sachdev,
Nucl.Phys. {\bf B316}, 609(1989).

\bibitem{Berry-3}N.Read and S.Sachdev,
Phys. Rev.B {\bf 42}, 4568(1990).

\bibitem{CS}The effects of the Berry phase were studied by 
doubled Chern-Simons theories rather in details by 
C.Xu and S.Sachdev, Phys.Rev.B {\bf 79}, 064405(2009).

\bibitem{CPN-1}I. Ya. Aref\'eva and S.I. Azakov,\\
Nucl.Phys. {\bf B162}, 298(1980).

\bibitem{CPN-2}S.Takashima, I.Ichinose, and T.Matsui, \\
Phys. Rev.B {\bf 72}, 075112(2005).

\bibitem{CPN-3}For numerical study on the noncompact CP$^1$ U(1) \\
gauge model, see O.I.Motrunich and A.Vishwanath, \\
Phys. Rev.B {\bf 70}, 075104(2004).

\bibitem{Yoshi}D.Yoshioka, G.Arakawa, I.Ichinose, and T.Matsui, \\
Phys. Rev.B {\bf 70}, 174407(2004).

\bibitem{Janke1}S. Wenzel and W. Janke,
Phys. Rev.B {\bf 79}, 014410(2009).

\bibitem{nonlocal}G.Arakawa, I.Ichinose, T.Matsui, and K.Sakakibara, \\
Phys. Rev. Lett.{\bf 94}, 211601(2005).

\bibitem{nonlocal2}
G.Arakawa, I.Ichinose, T.Matsui, K.Sakakibara, and \\ S.Takashima, 
Nucl.Phys. {\bf B732}[FS], 401(2006).

\bibitem{FNLambda}Corresponding to the spin model (\ref{Hspin}),
the only single component $\Lambda_{\alpha=1}$ appears in the
effective field theory. However, here we also introduce $\Lambda_{\alpha=2}$
for general consideration.

\bibitem{FNgauge}More precisely, in order to introduce the new complex field 
$v(x)$ through Eq.(\ref{zv}), we have to fix the gauge. 
The most convenient one is, e.g., $\Lambda_1=$real for the case
$\langle \Lambda_\alpha\rangle=\langle \Lambda_1\rangle\delta_{\alpha 1}$.

\bibitem{FNU1}From Eq.(\ref{zv}), it is obvious that 
a global U(1) phase rotation of $v(x)$ corresponds to a spatial translation.

\bibitem{massiveCPN}P.Azaria, P.Lecheminant, and D.Mouhanna, \\
Nucl. Phys. {\bf B455}, 648(1995).

\bibitem{vison}T.Senthil and M.P.A.Fisher, \\
Phys. Rev.B {\bf 63}, 134521(2001).

\bibitem{CSS}A.V.Chubkov, S.Sachdev, and T.Senthil,\\
Nucl. Phys. {\bf B426}, 601(1994).

\bibitem{nematic}P.E.Lammert, D.S.Rokhsar, and J.Toner, \\
Phys. Rev. Lett.{\bf 70}, 1650(1993).

\bibitem{MC}N.Metropolis, A.W.Rosenbluth, M.N.Rosenbluth, \\
A.M.Teller, and E.Teller,
J. Chem. Phys.{\bf 21}, 1087(1953).

\bibitem{Inst-1}T.A.DeGrand and D.Toussaint, \\
Phys. Rev.{\bf D22}, 2478(1980).

\bibitem{MHM}A.M.Ferrenberg and R.H.Swendsen, \\
Phys.Rev.Lett. {\bf 63}, 1195(1989).

\bibitem{dual}See for example, C.Itzykson and J.-M. Drouffe,
Chap.6 of {\it ``Statistical field theory"}(Cambridge University Press,
1989).

\bibitem{NKIM}K.Nakane, T.Kamijo, I.Ichinose, and T.Matsui,
work in progress.

\bibitem{FNBerry1}The Berry phase may move the location of 
the phase transition point and also change the order of 
the phase transition from second to first\cite{first-1,first-2}.

\bibitem{first-1}A.B.Kukulov, N.V.Prokofev, B.V.Svistunov, and \\
M.Troyer,
Ann.Phys.{\bf 321}, 1602(2006).

\bibitem{first-2}S.Kragset, E.Sm\o rgrav, J.Hove, F.S.Nogueira, and \\
A.Sudb\o, Phys.Rev.Lett. {\bf 97}, 247201.

\bibitem{FNBerry2}One may think that by the condensation of $\vec{\Lambda}$,
U(1) gauge field reduces to $Z_2$ and then the Berry phase becomes 
ineffective to instanton because discussion of instanton suppression 
in Refs.\cite{Berry-2,Berry-3} is not directly applicable to the 
$Z_2$ gauge theory.
However the instanton density can be defined without any ambiguity in the
$Z_2$ gauge theory if we use the spacetime lattice regularization.
It is an interesting problem to see if the instanton suppression argument
in the U(1) gauge theory survives in the $Z_2$ theory or not.




\end{references}
\end{document}